\documentclass[12pt]{article}

\usepackage{amsfonts}
\usepackage{amsmath}
\usepackage{esint}
\usepackage[margin=1in]{geometry}
\usepackage{graphicx}
\usepackage{epstopdf}
\usepackage{amssymb}
\usepackage{subcaption}
\usepackage{float}
\usepackage{caption}
\usepackage{color}
\usepackage{array}
\usepackage{url}
\usepackage{xcolor}
\usepackage{multicol}
\usepackage{hyperref}
\usepackage{wrapfig}

\hypersetup{
  colorlinks = true, %
  urlcolor   = black, %
  linkcolor  = black, %
  citecolor  = black %
}

\definecolor{myorange}{rgb}{0.8500 0.3250 0.0980}
\definecolor{myyellow}{rgb}{0.9290 0.6940 0.1250}
\definecolor{mygreen}{rgb}{0.4660 0.6740 0.1880}
\definecolor{mypurple}{rgb}{0.4940 0.1840 0.5560}
\definecolor{mybrown}{rgb}{0.53 0.26 0.12}
\definecolor{myred}{rgb}{1 0 0}
\definecolor{mycyan}{rgb}{0 1 1}

\numberwithin{equation}{section}

\title{Data-driven modeling and prediction of microglial cell dynamics in the ischemic penumbra} 
\author{\large Sara Amato$^{1}$, Andrea Arnold$^{1,2,*}$}
\date{}

\begin{document}
\maketitle

\small %

\centerline{$^1$ Bioinformatics \& Computational Biology Program, Worcester Polytechnic Institute, Worcester, MA, USA}
\vspace{.1cm}

\centerline{$^2$ Department of Mathematical Sciences, Worcester Polytechnic Institute, Worcester, MA, USA}
\vspace{.2cm}

\centerline{* Corresponding author: anarnold@wpi.edu, ORCID \href{https://orcid.org/0000-0003-3003-882X}{0000-0003-3003-882X}}

\bigskip

\begin{abstract}
Neuroinflammation immediately follows the onset of ischemic stroke. During this process, microglial cells are activated in and recruited to the tissue surrounding the irreversibly injured infarct core, referred to as the penumbra. Microglial cells can be activated into two distinct phenotypes; however, the dynamics between the detrimental M1 phenotype and beneficial M2 phenotype are not fully understood. 
Using phenotype-specific cell count data obtained from experimental studies on middle cerebral artery occlusion-induced stroke in mice, we employ sparsity-promoting system identification techniques combined with Bayesian statistical methods for uncertainty quantification to generate continuous and discrete-time predictive models of the M1 and M2 microglial cell dynamics.
The resulting data-driven models include constant and linear terms but do not include nonlinear interactions between the cells. Results emphasize an initial M2 dominance followed by a takeover of M1 cells, capture potential long-term dynamics of microglial cells, and suggest a persistent inflammatory response.\\

\end{abstract}

\normalsize

\section{Introduction}

Microglial cells are key components in the neuroinflammatory response following ischemic stroke, as they play a significant role in maintaining the health of the central nervous system and become activated after acute brain injury. Since the different phenotypes of these cells can be both beneficial (M2 phenotype) and detrimental (M1 phenotype) to stroke outcome, microglia are relevant clinical targets for intervention; however, the dynamics between the microglial cell phenotypes are not fully understood \cite{Boche2013}. The overarching goal of this study is to use experimental data and computational modeling techniques to generate data-driven predictive models of microglial cells after stroke onset to help inform our biological understanding of these cells and make inferences about their behavior over time.

In this paper, we focus on ischemic stroke as a result of middle cerebral artery occlusion (MCAO) \cite{Moulin1985, Feske2021, Uzdensky2019}. Following ischemic stroke due to MCAO, the ischemic core emerges as an area of irreversible damage. Throughout this work we refer to the vulnerable tissue surrounding the ischemic core as the penumbra. Since the penumbra can be salvaged with timely intervention, it is the most clinically-relevant target for treatment \cite{Feske2021, Uzdensky2019}. Clinicians are currently researching potential new therapeutic strategies using neuroprotectants, which aim to alter the cellular environment of the penumbra by promoting M2 microglial cell activation and suppressing M1 activation \cite{Lee2014, Guruswamy2017, Ginsberg2008, Zhao2017, Yenari2010, Yang2021}.

Microglial cell activation occurs in the penumbra immediately after the onset of ischemia \cite{Ma2017}. M2 microglial cells have been shown to dominate at the early stages of inflammation, whereas M1 microglial cells activate more slowly and then become the dominant phenotype for the remainder of the neuroinflammatory process \cite{Ma2017, Hu2012}. Once activated, microglial cells may be able to shift between phenotypes; however, the switching from the M2 to M1 phenotype has been cited as an area for further research \cite{Boche2013}. Additionally, the duration of the neuroinflammatory process is not well understood. Most MCAO ischemic stroke studies only measure microglial cell counts within a period of 14 days \cite{Hu2012, Li2018, Ma2020,Wang2017,Xu2021, Li2023, Yang2017}. Therefore, after the initial two weeks post ischemic stroke, there is not enough evidence to clearly state the dynamics of microglial cells.

Knowledge of the cellular interactions over time, both in the short-term and long-term, is crucial in understanding the neuroinflammatory process and how neuroprotective substances could work in the treatment of ischemic stroke. Computational modeling approaches have the potential to contribute to this understanding by way of building predictive models based on observed data. 
While deterministic models (e.g., using differential equations) can be designed to follow observed trends and/or fit to experimental data, formulating the model terms typically requires making assumptions based on reasonable knowledge of the underlying mechanisms of the biological system of interest \cite{Murray_book, Calvetti_book}. When less is known about the system dynamics, data-driven modeling approaches offer an alternative to formulating the model equations by relying primarily on the observed data. This often involves using tools from machine learning, such as neural networks and dictionary learning techniques, to approximate the system dynamics and make forecast predictions \cite{NODE, Rackauckas2021UDE, Brunton2016}. 
In this work, we contribute two new data-driven mathematical models that focus on the microglial response to MCAO-induced ischemic stroke. Both approaches use M1 and M2 microglial cell data obtained from experimental studies and sparsity-promoting system identification methods based on modifications of the sparse identification of nonlinear dynamics (SINDy) algorithm combined with Bayesian statistical techniques for uncertainty quantification.

\subsection{Paper Contributions and Organization}

The main contributions of this work include the following: 
\begin{itemize}
    \item We contribute two new data-driven, sparse models of microglial cell dynamics post ischemic stroke: (i) a continuous-time dynamical system of the form in \eqref{eq: ODE} with stochastic coefficients, and (ii) a discrete-time model of the form in \eqref{eq: DTSINDy} prescribing the mean of a statistical model formulation, both using experimental data of M1 and M2 cell counts in the penumbra area of the ischemic brain.
    \item We develop a numerical framework which combines SINDy-based model identification with Bayesian statistical techniques to obtain robust data-driven models appropriate for forecast predictions and uncertainty quantification.
    \item We use the resulting predictive models to help inform our understanding of microglial cell interactions after the onset of MCAO-induced stroke and make inferences about the microglial cell behavior over time.
\end{itemize}
In deriving these models, we detail two different approaches which combine machine learning (via SINDy and ESINDy) with computational techniques from statistical inverse problems, sensitivity analysis, parameter estimation, and uncertainty quantification to make data-driven predictions of microglial cell counts throughout the neuroinflammatory process.

Since our goal in developing data-driven models is not to assume a known mechanistic form but rather to discover the governing equations from observed data, we consider two different approaches for determining the sparse model and quantifying uncertainty in the forecast predictions. 
Our first approach uses SINDy to determine a sparse, continuous-time system of ordinary differential equations (ODEs) modeling the microglial cell dynamics:
\begin{equation}
    \dot{x}(t) = f(x(t); \theta)
    \label{eq: ODE}
\end{equation}
where the right-hand side function $f:\mathbb{R}^2\rightarrow\mathbb{R}^2$ is unknown and depends on the coefficients $\theta$ representing the weights of the basis functions in the resulting sparse model. 
While the SINDy algorithm provides a point-estimate of $\theta$, we introduce stochasticity into the deterministic ODE model by treating these parameters as random variables and utilizing a robust parameter estimation procedure, based on global sensitivity analysis and Markov Chain Monte Carlo (MCMC) sampling, to estimate a probability distribution for the most influential parameters and make forecast predictions with forward propagation of uncertainty. 
In the second method, we use Ensemble-SINDy (ESINDy) to identify the most likely Dynamic Bayesian Network (DBN) structure and formulate a sparse, discrete-time model for the microglial cell dynamics:
\begin{equation}
    {x}_{t+1} = {g}({x}_t; \theta) \label{eq: DTSINDy}
\end{equation}
where the right-hand side function $g: \mathbb{R}^2\rightarrow\mathbb{R}^2$ is unknown and again depends on some model coefficients $\theta$. 
Since DBN theory provides a natural statistical framework to perform uncertainty quantification, we need not apply additional parameter estimation techniques to this model. 
We identify both models using data sets inspired by experimentally-observed M1 and M2 microglial cell counts in the penumbra from murine MCAO-induced stroke models in adult male mice and a set of candidate basis functions, analyzing forecast predictions from both models to provide insights on microglial cell dynamics over time.

Previous models using Bayesian networks and DBNs have explored interactions between neuroinflammatory components including cytokines, chemokines, and microglia in traumatic brain injury (TBI) and ischemic stroke \cite{Zeng2013, Lewis2019, Azhar2021, Abboud2016}. However, to our knowledge, DBNs have not been used to study the interactions between M1 and M2 microglial cell phenotypes prior to this work. Previous ODE models involving the two microglial cell phenotypes consider TBI, amyotrophic lateral sclerosis, hemorrhagic shock, Alzheimer's disease, general neuroinflammation, and ischemic stroke \cite{Alqarni2021, Hao2016, Shao2013, Vaughan2018, Amato2021, Amato2024a}. In particular, the work in Amato and Arnold (2025)~\cite{Amato2024a} presents a deterministic ODE model with stochastic coefficients for predicting microglial cell dynamics post ischemic stroke; while the model parameters were estimated from experimental data, the proposed dynamical system was not learned directly from the data. While researchers have used the SINDy algorithm to discover governing equations driving biological data sets such as COVID-19 data, yeast glycolysis data, and gene expressions \cite{Jiang2021, Mangan2016, Sandoz2023}, this approach has not previously been used to derive models of neuroinflammation. 
Our contributions in this work focus on developing new data-driven models of microglial cells to explore the interactions between the phenotypes as well as to predict their dynamics over time by combining existing machine learning and Bayesian techniques in a novel way for this application.

The paper is organized as follows. 
Section~\ref{sec: Methods} introduces the two proposed data-driven modeling methods: SINDy+MCMC, which uses SINDy to identify a sparse, continuous-time ODE model combined with classic techniques from sensitivity analysis and parameter estimation to perform uncertainty quantification; and ESINDy+DBN, which uses ESINDy to identify the most likely DBN structure and formulate a sparse, discrete-time model within a natural statistical framework to perform uncertainty quantification.
Section~\ref{sec: Results} presents the resulting models obtained by applying these methods in conjunction with the experimental data of microglial cell counts, including an analysis of the biological implications of the main results. 
Section~\ref{sec: Discussion} provides a discussion of the results and future work, and Section~\ref{sec: Conclusion} concludes the paper.

\section{Materials and Methods}
\label{sec: Methods}

This section introduces two methods used to derive data-driven models of the microglial cell dynamics. 
Both methods are based on modifications of the SINDy algorithm combined with Bayesian statistical techniques for uncertainty quantification. Here we provide an overview of these computational techniques while focusing on the specific application to microglial cell dynamics during the neuroinflammatory process post ischemic stroke.
We encourage interested readers to refer to the works of Brunton et al. (2016) \cite{Brunton2016} and Fasel et al. (2022) \cite{Fasel2022} for more details on SINDy and ESINDy, respectively; Sobol (1993) \cite{Sobol1993}, Smith (2013) \cite{Smith2013}, and Gamboa et al. (2013) \cite{Gamboa2013} for sensitivity analysis and Generalized Sobol Sensitivity (GSS) analysis; Metropolis et al. (1953) \cite{Metropolis1953}, Hastings (1970) \cite{Hastings1970}, Haario (2001) \cite{Haario2001}, Haario (2006) \cite{Haario2006}, Robert and Casella (1999) \cite{Robert2004}, Cotter et al. (2013) \cite{Cotter2013}, and Calvetti and Somersalo (2023) \cite{Calvetti2023} for MCMC methods; and Koller and Friedman (2009) \cite{Koller2009} for more details on DBNs.

\subsection{SINDy+MCMC Method for Continuous-Time Model}
\label{sec: MethodSINDyDE}

The goal of our first approach is to determine a continuous-time predictive model for the microglial cell dynamics.
The following method, which we refer to as SINDy+MCMC, has three main steps. First, we use the SINDy algorithm to obtain a sparse model describing the M1 and M2 microglial cell dynamics over time directly from experimentally-observed time series data. Second, to quantify uncertainty, we use GSS analysis to rank the resulting SINDy model parameters according to their influence on a combined output of M1 and M2 microglial cells and estimate the three most sensitive parameters using MCMC-based parameter estimation. Third, we utilize the point estimates from SINDy and random samples from the posterior distributions of the MCMC-estimated parameters to provide forecast predictions with forward propagation of uncertainty and analyze results through a biological lens. 
Amato and Arnold (2024) \cite{AmatoArnold2024_proc} presents preliminary results for this approach.

\subsubsection{Data}

\begin{figure}[t!]
\centerline{\includegraphics[width = 0.5\textwidth]{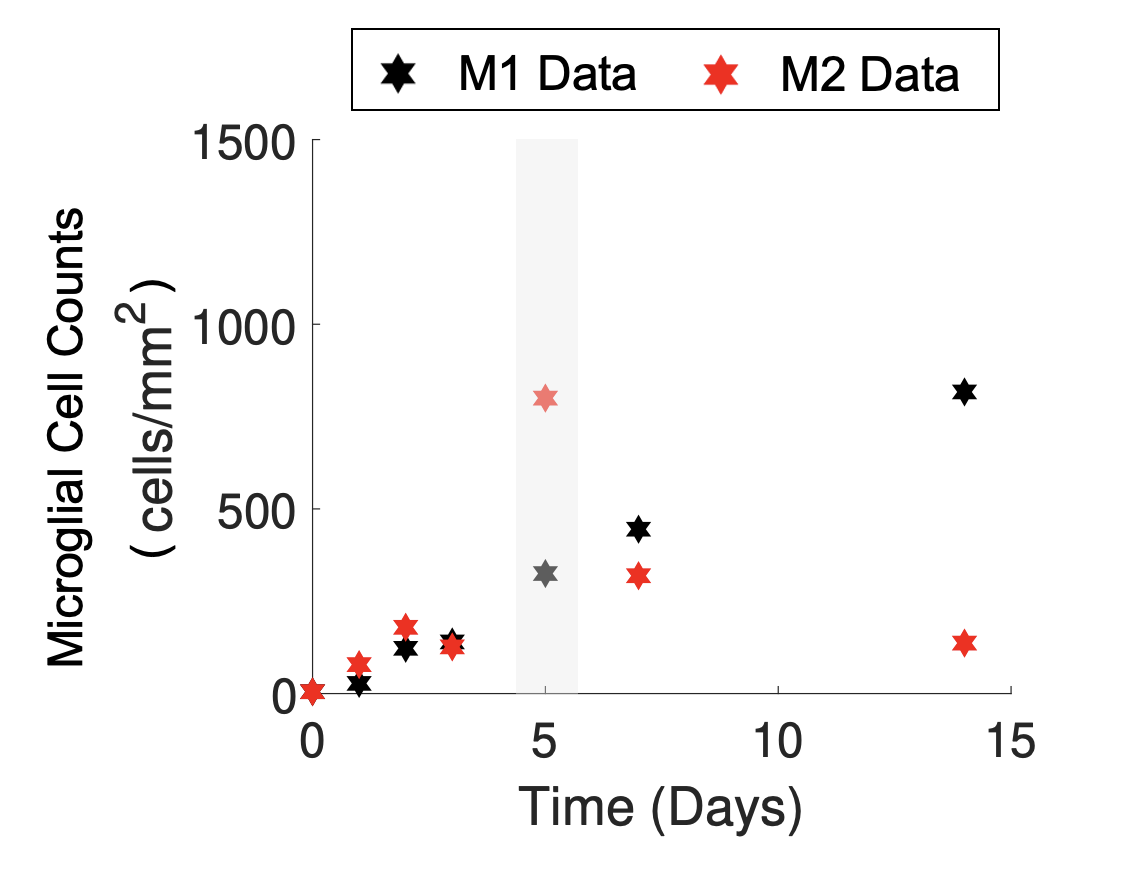}}
\caption{Time series data for M1 and M2 microglial cell counts compiled from experimental studies of MCAO-induced stroke in adult male mice. The potential outlier data at Day 5 are highlighted in grey. Note that we omit these observations during the SINDy step but include them for sensitivity analysis and parameter estimation.}
\label{fig: Data}
\end{figure}

The experimental data set used within the SINDy+MCMC method is a compiled time series of observed M1 and M2 microglial cell counts obtained from experimental studies using adult male mice models of permanent and transient MCAO \cite{Hu2012, Li2018, Ma2020, Wang2017, Xu2021, Li2023, Suenaga2015, Yang2017}. The data points are presented as averaged microglial cell counts from the penumbra area of the ischemic brain. Table~\ref{table: papers} and Figure~\ref{fig: Data} summarize the data; Amato and Arnold (2025) \cite{Amato2024a} and Amato (2024) \cite{AmatoThesis} provide a more thorough description of the experimental studies and data compilation. Within the SINDy step, we omit a potential outlier at Day 5 so that learned dynamics are not dependent on this point; however, within the robust parameter estimation step, we include the data at Day 5 to utilize the full set of available data.

\begin{table}[t!] 
\centering
\renewcommand{\arraystretch}{1.2}
\begin{tabular}{| p{.3\linewidth} | p{.3\linewidth} | p{.3\linewidth} |} 
 \hline
Reference & M1 $\&$ M2 Markers & Measurement Times (Days)\\
 \hline\hline
Hu et al. (2012)~\cite{Hu2012} & CD16/CD32, CD206 & 0, 1, 3, 5, 7, 14\\
    \hline
Li et al. (2018)~\cite{Li2018} & CD16, CD206 & 1, 3, 7\\
    \hline
Ma et al. (2020)~\cite{Ma2020} & CD16/CD32, CD206/Iba1  &  2\\
    \hline
Wang et al. (2017)~\cite{Wang2017} & CD16, CD206 & 14\\
    \hline
Xu et al. (2021)~\cite{Xu2021} & CD16/CD32, CD206 & 3\\
    \hline
Li et al. (2023)~\cite{Li2023} & CD16/CD32, CD206 & 2\\
    \hline
Suenaga et al. (2015)~\cite{Suenaga2015} & CD16/CD32, CD206 & 0, 1, 3, 7, 14\\
    \hline
Yang et al. (2017)~\cite{Yang2017} & CD16, CD206 & 3, 7\\
\hline 
\end{tabular}
\caption{Summary of the experimental studies used in compiling the data set shown in Figure~\ref{fig: Data}. 
The article reference, markers used to identify the microglial cell phenotypes, and the measurement days are provided for each study. 
M1 and M2 cell counts are averaged across studies to obtain a single set of time series data. 
More details are given in Amato and Arnold (2025) \cite{Amato2024a} and Amato (2024) \cite{AmatoThesis}.
}
\label{table: papers}
\end{table}

\subsubsection{SINDy}

We apply the SINDy algorithm \cite{Brunton2016} to determine a sparse representation of the mapping $f(\cdot)$ in \eqref{eq: ODE}. Briefly, we construct a candidate function library, $\Theta(X)$, implemented as a matrix where each column represents a potential candidate to be included in the functional form of $f(\cdot)$ evaluated at the discrete times of observed data, and use Sequential Threshold Least Squares (STLS) to solve for $\Xi$ in
\begin{equation}
    \dot{X} = \Theta(X) \Xi
\end{equation}
where $\dot{X} = [\dot{M1}(t) \ \ \dot{M2}(t)]$ and $\Xi$ is a matrix of coefficients multiplying the candidate functions. 
We use $M1(t)$ and $M2(t)$, the microglial cell time series data omitting Day 5, to construct $\Theta(X)$ and include a constant term, linear terms, and quadratic nonlinearities to represent dynamics that could potentially occur between microglial cells:
\begin{equation}
      \Theta(X) = \begin{bmatrix}
1 & M1(t_1) & M2(t_1) & M1(t_1)^2 & M2(t_1)^2 & M1(t_1)M2(t_1) \\
\vdots & \vdots & \vdots & \vdots & \vdots & \vdots\\
1 & M1(t_{end}) & M2(t_{end}) & M1(t_{end})^2 & M2(t_{end})^2 & M1(t_{end})M2({t_{end}})
\end{bmatrix} .  \label{eq: ThetaX}
\end{equation}
Note that the candidate functions in $\Theta(X)$ could include higher-order polynomial terms, trigonometric functions, or exponential functions; however, the number of candidate functions should not exceed the number of data points \cite{Brunton2016}. When including terms like these in initial simulations along with the candidate functions in \eqref{eq: ThetaX}, model fits were similar to those in the results section; however, the models were not sparse. Additionally, the biological relevance of such terms to the problem of interest is unclear. For these reasons, we do not include additional candidate functions in \eqref{eq: ThetaX} for this application.

\begin{figure}[t!]
\centering
\includegraphics[width = 0.7\linewidth]{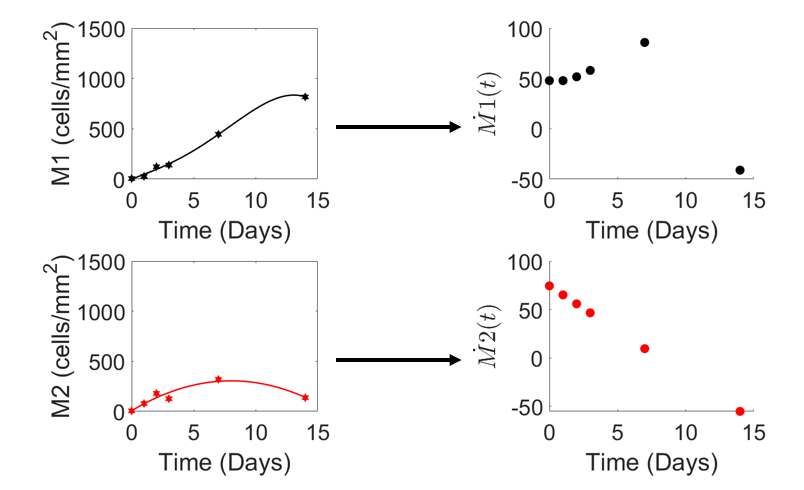}%
\caption{Derivative approximation used in the SINDy algorithm. We fit a fourth-order and second-order polynomial to the M1 and M2 microglial cell data, respectively, and analytically compute the derivatives at the discrete times of the observed data.
}
\label{fig: derivs}
\end{figure}

Prior works use a variety of methods to estimate derivatives within the SINDy method, including finite-difference methods, total-variation regularization, and analytic computation by fitting data to third-order polynomials \cite{Jiang2021, Mangan2016, Sandoz2022}. Finite-difference methods work well for approximating derivatives from non-noisy data; however, if the data are noisy, other methods should be considered. Total-variation regularization has been shown to work well for approximating derivatives from noisy data; however, it is often assumed that the noise has a known distribution \cite{Chartrand2011}. If data are noisy with an unknown noise distribution, one can fit the data to a polynomial and analytically estimate the derivative \cite{Sandoz2022}. We approximate derivative information for the M1 and M2 microglial cells using a fourth-order LS polynomial fit to the M1 microglial cell data and a second-order LS polynomial fit to the M2 microglial cell data and analytically compute the derivatives in each case as shown in Figure~\ref{fig: derivs}.

The STLS algorithm, an iterative parameter approximation technique, alternates between finding a LS solution and discarding parameter values smaller than a prescribed threshold parameter $\lambda$.
That is, starting with an initial guess for $\Xi$ using the LS solution, we solve  
\begin{equation}
    \min_{\Xi} ||\dot{X} - \Theta(X)\Xi||_2 + \lambda ||\Xi||_0
\end{equation}
where $||\cdot||_0$ denotes the number of nonzero entries in the vector. The process iterates by setting terms less than $\lambda$ equal to 0 and computing the LS solution of the remaining system until convergence, with the aim to balance error and sparsity in the resulting model.

\subsubsection{GSS Analysis and MCMC-Based Parameter Estimation}
Although the SINDy method yields point estimates for the parameters of the resulting ODE model, which can be used to propagate the model forward in time, we must use a more robust parameter estimation procedure to obtain a sense of uncertainty in our parameter estimates and forecast predictions. To this end, we utilize GSS analysis to determine the most influential parameters in our model, taking into account both the M1 and M2 microglial cell counts as outputs \cite{Gamboa2013}. 
We then apply the Metropolis-Hastings (MH) algorithm to estimate the three most sensitive parameters using the MCMC toolbox for MATLAB \cite{mcmcstat}. 
Note that the MCMC toolbox software implements different MCMC sampling approaches, including MH and its variants. 
While alternative approaches such as the Delayed Rejection Adaptive Metropolis (DRAM) algorithm could be used in place of MH in this step, the results are similar when using MH and DRAM as the sampling methods for this problem.

\subsubsection{Forecast Predictions with Forward Propagation of Uncertainty}
The MCMC sampling procedure results in a posterior sample for each of the estimated model parameters. Therefore, there are different combinations of parameter values (in addition to the mean estimates) that can be used from the posterior to simulate the forward model. Additionally, since experimental data are only collected over a two-week period, it is of interest to simulate predicted microglial cell counts beyond the 14 days of observation.  In an effort to quantify uncertainty in the model outputs and propagate this uncertainty forward in time with model predictions, we draw $N$ random samples from the estimated parameter posteriors and solve the forward model over the time interval $[0, 50]$ days with each of these $N$ parameter sets. After computing the forward model simulations, there are $N$ predicted values that we use to calculate a mean and standard deviation at discretized time points over the specified time interval.

\subsection{ESINDy+DBN Method for Discrete-Time Model}
\label{sec: MethodESINDyDBN}

The goal of our second approach is to determine a discrete-time predictive model for the microglial cell dynamics.
The following method, which we refer to as ESINDy+DBN, combines the theory of DBN statistical modeling with ESINDy and has two main steps. First, we use the ESINDy algorithm to infer topologies of DBNs for microglial cells directly from the time series data, as well as to determine the weights of the connections and to quantify the likelihood of possible network structures. Second, we use the inferred interactions and estimated weights, along with DBN theory, to model the microglial cell dynamics over time and determine uncertainty in our estimates. For this approach, note that we do not need to use MCMC to get parameter distributions, since learning uncertainty is involved in the model identification process for DBNs.

\subsubsection{Data} \label{subsec: ESINDy_proxydata}

\begin{figure}[t!]
\centerline{\includegraphics[width = 0.44\textwidth]{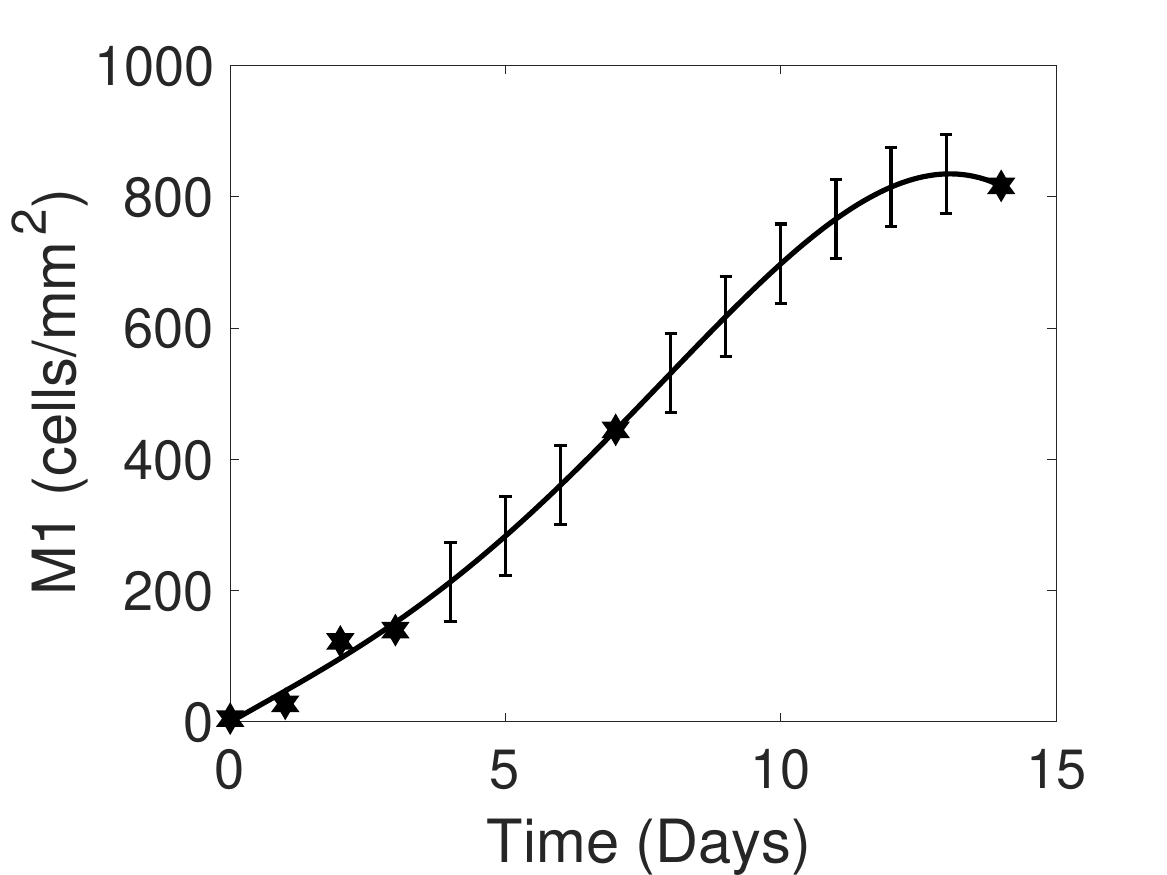}\includegraphics[width = 0.44\textwidth]{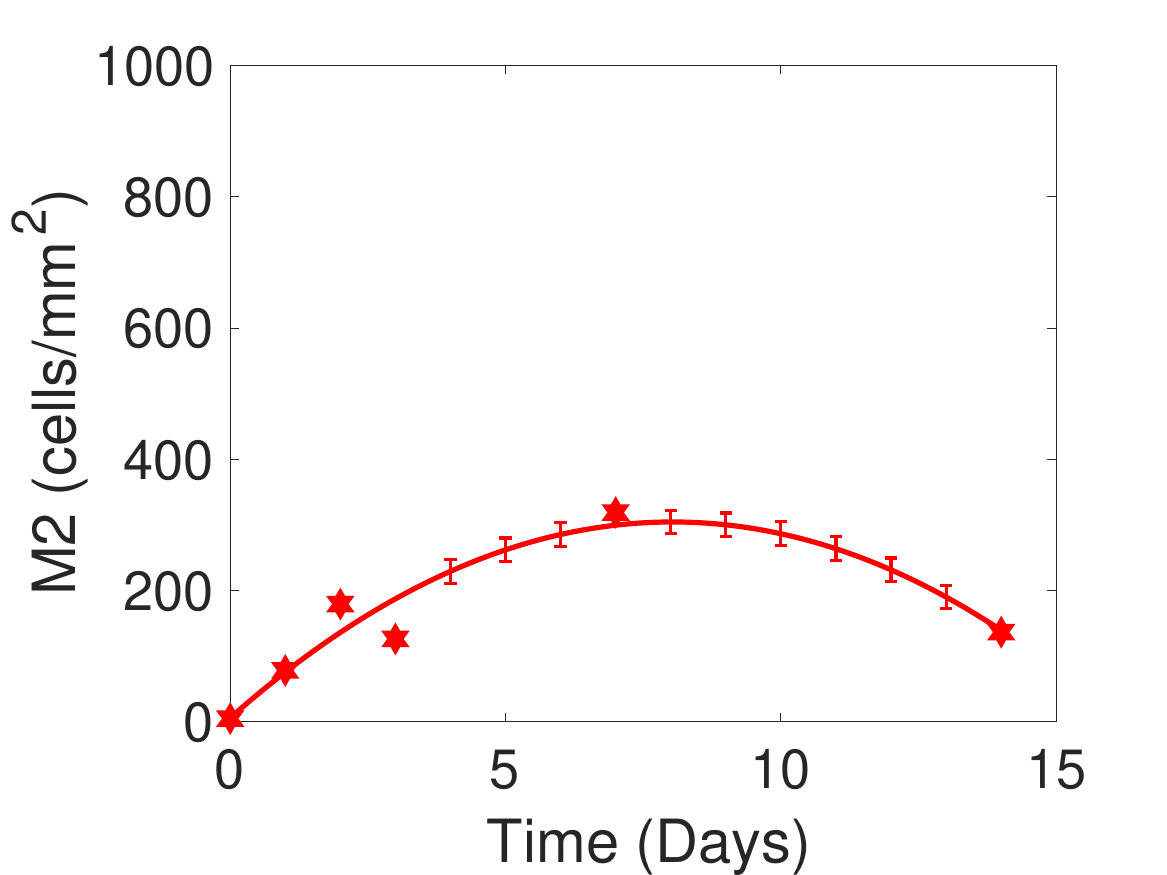}}
    \caption{Proxy data generation for the ESINDy+DBN method. We fit a fourth-order polynomial (black curve) and a quadratic polynomial (red curve) to the M1 and M2 microglial cell count data, respectively. To generate the proxy data, we add zero-mean Gaussian noise with standard deviation equivalent to $10\%$ the standard deviation of the M1 and M2 curves to the approximate cell counts from the polynomial fits on days where no experimental measurements are given. Black and red stars plot the averaged microglial cell experimental data (used within each ESINDy run), whereas black and red error bars show 2 times the standard deviation of Gaussian noise around the mean of each curve at each time point where proxy data were drawn.}
    \label{fig: DataESINDy}
\end{figure}

To effectively learn and provide a correct interpretation of DBNs, we must use frequent-in-time equispaced data \cite{Azhar2021, Huang2007, Kim2003}. We start by fitting a fourth-order polynomial to the compiled M1 experimental data and a quadratic polynomial to the compiled M2 experimental data in Figure~\ref{fig: Data}, with the potential outlier at Day 5 omitted, as shown in Figure~\ref{fig: derivs}. To increase the time frequency of observed data, we augment the experimental data set with fictitious cell count observations at the days with no recorded measurements (i.e., at Days 4, 5, 6, 8, 9, 10, 11, 12, and 13), which we generate by adding Gaussian noise to the fitted M1 and M2 polynomial curves. We set the standard deviations of the Gaussian noise terms equivalent to $10\%$ of the standard deviations of the fitted M1 and M2 curves, respectively, to more closely simulate experimental data. 
Figure~\ref{fig: DataESINDy} depicts the proxy data generation, where the error bars show 2 times the standard deviation of Gaussian noise around the mean of each curve.
We apply the ESINDy+DBN method to multiple proxy data sets to account for different realizations of noise in the simulated measurements and compute the average parameter values for each coefficient included in all models.

\subsubsection{ESINDy}
To determine a sparse representation of $g(\cdot)$ in \eqref{eq: DTSINDy} for each proxy data set, we use the discrete-time formulation of SINDy paired with the library ESINDy algorithm \cite{Brunton2016, Fasel2022}. 
In this case, we use ESINDy instead of SINDy to more closely match with DBN theory, placing a strong emphasis on quantifying model uncertainty.
For the discrete-time formulation of SINDy, we construct a candidate function library, $\Theta(X_t)$, a matrix where each column represents a potential candidate to be included in the function form of $g(\cdot)$, and use STLS to solve for $\Xi$ in
\begin{equation}
    X_{t+1} = \Theta(X_t) \Xi
\end{equation}
where
\begin{equation}
X_{t+1} = \begin{bmatrix}
M1_{t_2} & M2_{t_2} \\
\vdots & \vdots \\
{M1}_{t_{end}} & {M2}_{t_{end}}
\end{bmatrix}
\end{equation}
and $\Xi$ is a matrix of coefficients multiplying the candidate functions. We use the simulated equispaced data in Figure~\ref{fig: DataESINDy} to construct $\Theta(X_t)$ and similarly include a constant term, linear terms, and quadratic nonlinearities to represent dynamics that could potentially occur between microglial cells:
\begin{equation}
     \Theta(X_t) = \begin{bmatrix}
1 & M1_{t_1} & M2_{t_1} & M1_{t_1}^2 & M2_{t_1}^2 & M1_{t_1}M2_{t_1} \\
\vdots & \vdots & \vdots & \vdots & \vdots & \vdots\\
1 & M1_{t_{end-1}} & M2_{t_{end-1}} & M1_{t_{end-1}}^2 & M2_{t_{end-1}}^2 & M1_{t_{end-1}}M2_{{t_{end-1}}}
\end{bmatrix} .
\end{equation}
To perform ESINDy along with this discrete-time formulation, for a specified number of ensembles $S$, at each iteration, we randomly sample a subset of $h$ candidate functions from $\Theta(X_t)$, which we denote as $\Theta_s(X_t)$, and solve
\begin{equation}
    X_{t+1} = \Theta_s(X_t) \Xi_s
\end{equation}
using STLS. 
Note that for each $s \in (1, \ldots, S)$, $\Theta_s(X_t)$ is an $(N-1) \times h$ matrix and $X_{t+1}$ is an $(N-1) \times h$ matrix where $N$ is the number of time points considered (in this case, $N = 15$). Then $\Xi_s$ is a $h \times 2$ matrix of coefficients. In particular, $\Xi_s$ is the set of coefficients corresponding to the respective candidate functions in $\Theta_s(X_t)$.
When all ensembles are generated, we have $S$ sets of possible coefficients. We group together iterations that yield the same model and compute the likelihood of each model by counting the number of times it appeared and dividing by the total number of iterations.

\subsubsection{Dynamic Bayesian Networks and Forecast Predictions with Uncertainty}
\label{sec: DBN}

Results of the ESINDy algorithm provide a natural framework for constructing DBNs and greatly simplify the DBN structure learning process. Additionally, DBN theory provides a natural statistical framework to perform uncertainty quantification. Therefore, combining these two methods streamlines the process of identifying a discrete-time predictive model of the M1 and M2 microglial cell dynamics.

\begin{figure}[t!]
\centering
\includegraphics[width = 0.65\linewidth]{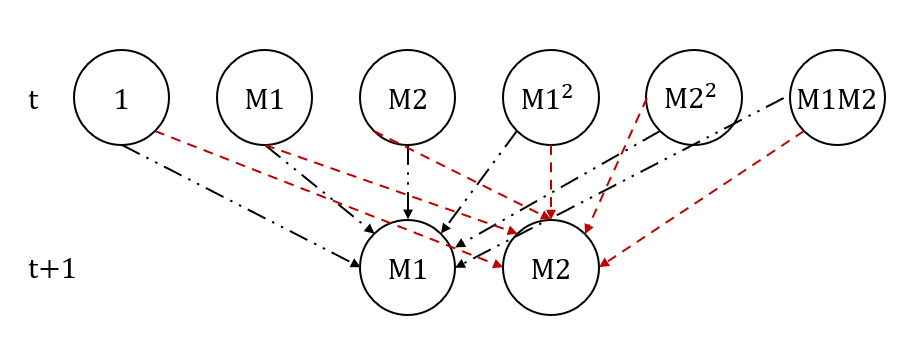}
    \normalsize{
    \begin{eqnarray}
    p(M1_{t+1}|Pa(M1)_t) &=& \mathcal{N}(\Theta_s(X_t) \Xi_s(1, :), \sigma^2_{M1_{t+1}}) \label{eq: DBNmodel5M1prob1} \\ [0.2cm]
    p(M2_{t+1}|Pa(M2)_t) &=& \mathcal{N}(\Theta_s(X_t)\Xi_s(2,:), \sigma^2_{M2_{t+1}}) \label{eq: DBNmodel5M2prob1}
\end{eqnarray}}
\caption{{Top:} DBN for M1 and M2 microglial cells with all possible connections shown by dashed lines. The connections repeat over time and only depend on one time period prior. {Bottom:} DBN models used in this work. We assume that M1 and M2 are each normally distributed with mean equivalent to a linear combination of their respective parents and variance prescribed. Means are found using the ESINDy method.}
  \label{fig: DBNModel}
    \end{figure}

Briefly, DBNs are probabilistic graphical models where the states, in this case M1 and M2, evolve over time and are treated as normally distributed random variables with first-order Markovian connections, i.e., the interactions do not change over time and only depend on one time period prior.  
The variables from one time period prior that contribute to the value of M1 and M2 over time are called the parents of M1 and M2, respectively. Following DBN theory, we assume that M1 and M2 are normally distributed with mean equivalent to a linear combination of their parents with weights equivalent to the strength of the connection and time-dependent variance. We summarize this information in Figure~\ref{fig: DBNModel}.

Learning a DBN from data involves finding the parents of each random variable, the weights of each connection, and the variances. Structure learning algorithms such as Reversible Jump MCMC have been used to learn the DBN topology with variances computed separately \cite{Huang2007, Abboud2016, Azhar2021}. In this step, we instead use results from ESINDy to construct the DBN and then compute the variance.
For each distinct $\Xi_s$ obtained from the ESINDy step, we can construct a DBN. Note that the candidate functions corresponding to the non-zero weights in the first and second columns of $\Xi_s$ are the parents of M1 and M2, respectively. We construct our topologies using this information and set the means of the normal distributions of M1 and M2 to be equivalent to $\Theta_s(X_t)\Xi_s(1,:)$ and $\Theta_s(X_t)\Xi_s(2,:)$, as shown in the bottom panel of Figure~\ref{fig: DBNModel}.

In the DBN literature, the variance parameter is set as a ratio where the numerator involves the mean of the random variable and the true data and the denominator involves a modification of the sample size. Inspired by this ratio and incorporating what we know about the averaged data, we introduce a standard deviation for each random variable that depends on the root mean square error between the mean estimate and averaged experimental data. The root mean square error (RMSE) allows us to incorporate information about the mean of M1 and M2 as well as to reincorporate the true data, since we are using proxy data to find the mean. Additionally, using the RMSE allows us to keep the spirit of the variance term in DBN literature.

\section{Results}
\label{sec: Results}
In this section, we present the resulting data-driven models obtained from the SINDy+MCMC method and the ESINDy+DBN method. 
We use MATLAB programming language (The MathWorks, Inc., Natick, MA) to implement the algorithms and compute numerical results.
While we are interested in the predictive power of our models, the microglial cell count data obtained from the experimental studies on MCAO mice are only collected over a time period of 14 days. Therefore, we use additional experimental data to gain information on the accuracy of our model forecast predictions. To this end, we analyze and evaluate forecast predictions based on information obtained from experimental literature on microglial cell counts following MCAO-induced ischemic stroke with measurements taken in the thalamus or ipsilateral brain regions and in neuroinflammation following TBI \cite{Suenaga2015, Rupalla1998, Bodhankar2015, Stubbe2013, Febinger2015, Younger2019, Donat2017}. 
In particular, we use the following three criteria: (i) the amount of M1 cells may still be significantly greater than the amount of M2 cells from Day 14 on; (ii) between Day 14 and Day 35 there may be some lingering inflammatory response; however, the form of this response is unknown (i.e., the number of cells may have reached a steady state,
may be oscillating, or may be increasing/decreasing slowly); and (iii) after Day 35, we may assume that the amount of M1 and M2 cells decreases towards their respective baseline value. Note that Amato (2024)~\cite{AmatoThesis} contains a more thorough summary of the papers cited above.

\subsection{SINDy+MCMC Method}

\begin{figure}[t!] 
\centerline{\includegraphics[width = 0.33\linewidth]{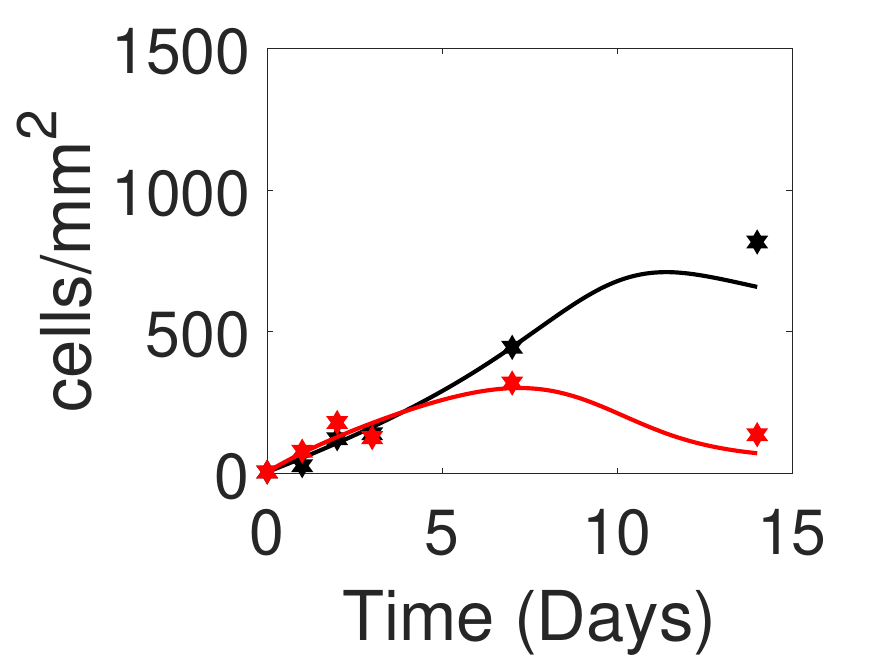}\includegraphics[width = 0.33\linewidth]{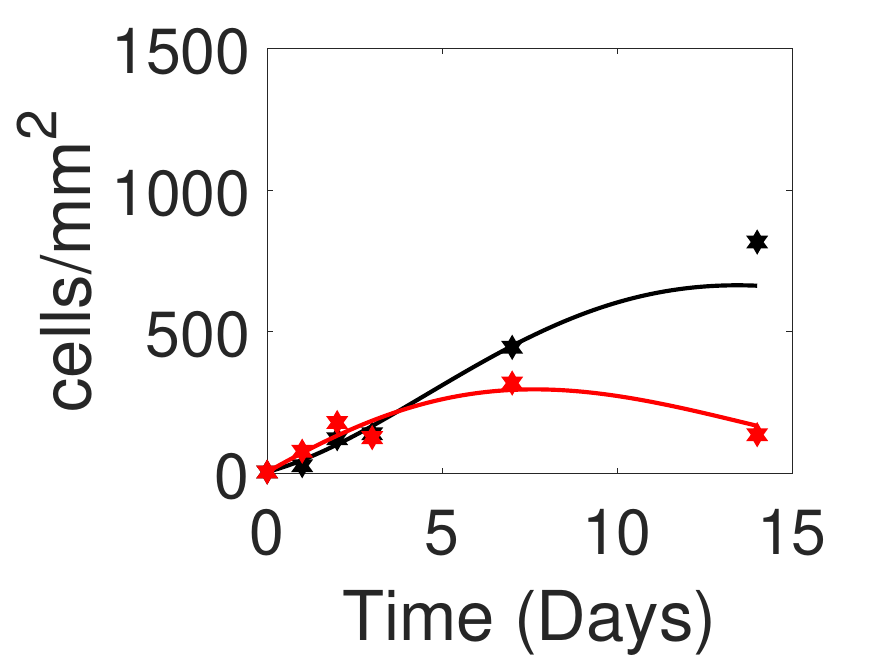}\includegraphics[width = 0.33\linewidth]{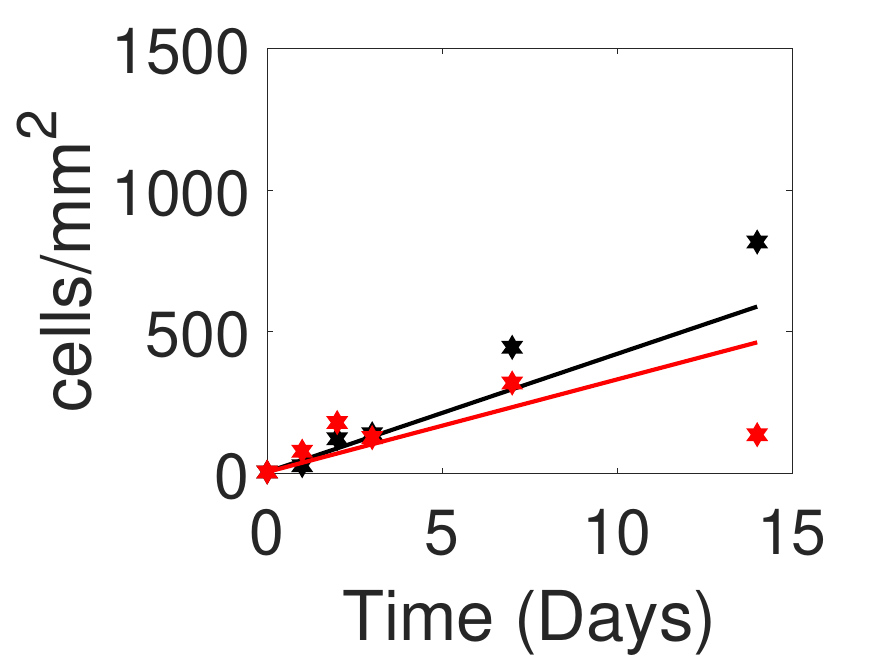}}
\caption{Time series approximations obtained from applying SINDy to the microglial cell count data in Figure~\ref{fig: Data} with various threshold parameters $\lambda$. From left to right: $\lambda = 0$, $\lambda = 0.01$, and $\lambda = 0.2$. }
\label{fig: Exp3SINDy}
\end{figure}

Figure~\ref{fig: Exp3SINDy} shows the results of applying the SINDy method to the M1 and M2 microglial cell data in  Figure~\ref{fig: Data} with different values of the STLS threshold parameter $\lambda$. Note that setting $\lambda = 0$ is equivalent to finding the LS solution, and all terms are active in describing the right-hand side dynamics. When $\lambda = 0.01$ only the linear candidate functions, i.e., the constant coefficient, M1 microglial cells, and M2 microglial cells, are active in describing the M1 and M2 microglial cell derivatives over time. Biologically, this may suggest that interaction terms are not needed to describe the microglial cell dynamics observed in this data set. 
When $\lambda = 0.2$, the resulting model only includes the constant terms.

\begin{figure}[t!]
\begin{minipage}{0.45 \textwidth} 
\centering
\includegraphics[width = 0.8 \linewidth]{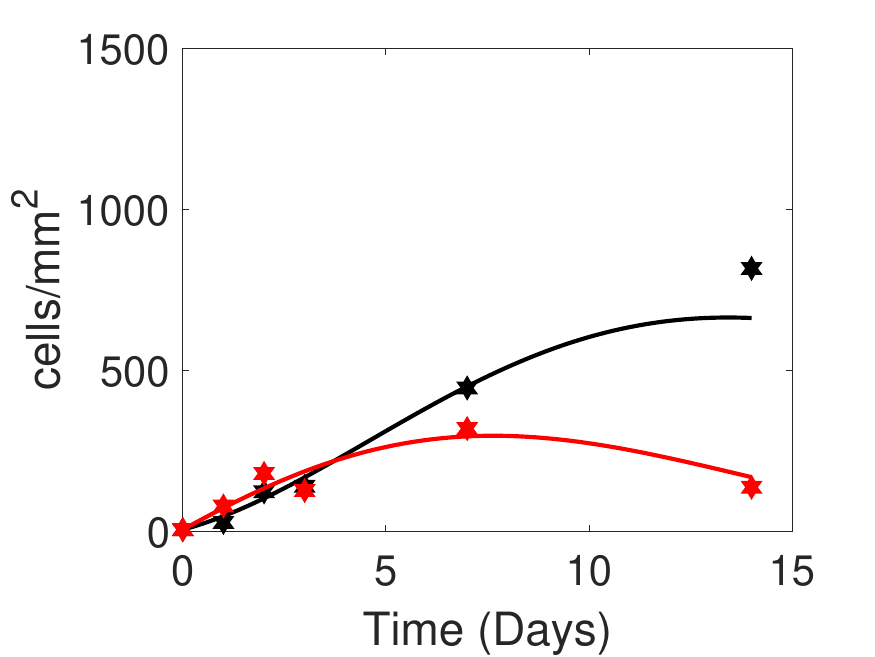}
    \end{minipage}
    \begin{minipage}{0.45 \textwidth}
    \centering
\includegraphics[width = 1\linewidth]{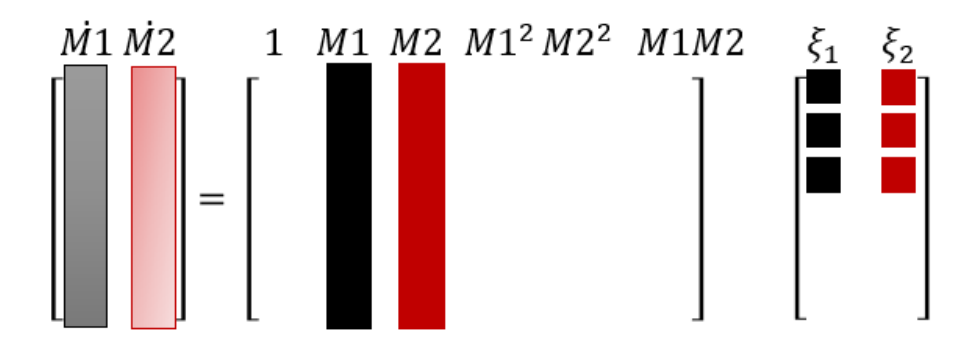}
\end{minipage}
\begin{eqnarray}
    \dot{M1}(t) &=& \underbrace{32.9650}_{\theta_1} + \underbrace{-0.1377}_{\theta_2} M1(t) + \underbrace{0.3157}_{\theta_3} M2(t) \label{eq: SINDyM1} \\
    \dot{M2}(t) &=& \underbrace{70.3415}_{\theta_4} + \underbrace{- 0.1569}_{\theta_5} M1(t) + \underbrace{0.0217}_{\theta_6} M2(t) \label{eq: SINDyM2} 
\end{eqnarray}
\caption{Resulting ODE model and simulated time series obtained from applying SINDy to the data summarized in Figure~\ref{fig: Data} with threshold parameter $\lambda = 0.01$. Note that only the linear terms are active within the resulting sparse model in \eqref{eq: SINDyM1} and \eqref{eq: SINDyM2}.}
\label{fig: DESINDyresults01}
\end{figure}

\begin{figure}[t!]
\centerline{\includegraphics[width = 0.5 \linewidth]{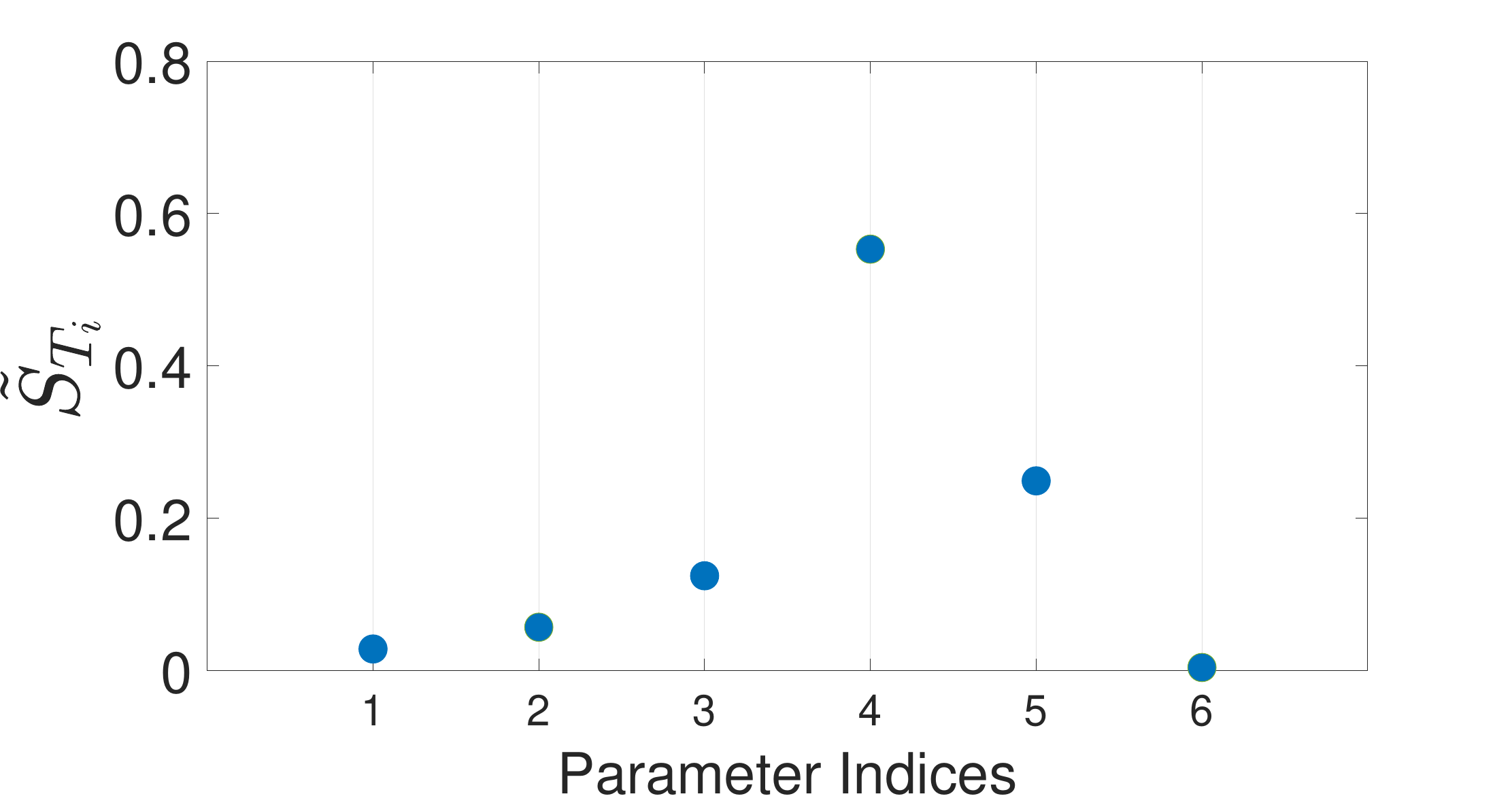}}
\caption{Results of GSS applied to the ODE model in \eqref{eq: SINDyM1}--\eqref{eq: SINDyM2}. The ranking of parameter values from most to least sensitive is: $\theta_4$, $\theta_5$, $\theta_3$, $\theta_2$, $\theta_1$, and $\theta_6$. }
\label{fig: DESINDyl01resultsSobol}
\end{figure}

\begin{figure}[t!]
\centerline{\includegraphics[width=.33\linewidth]{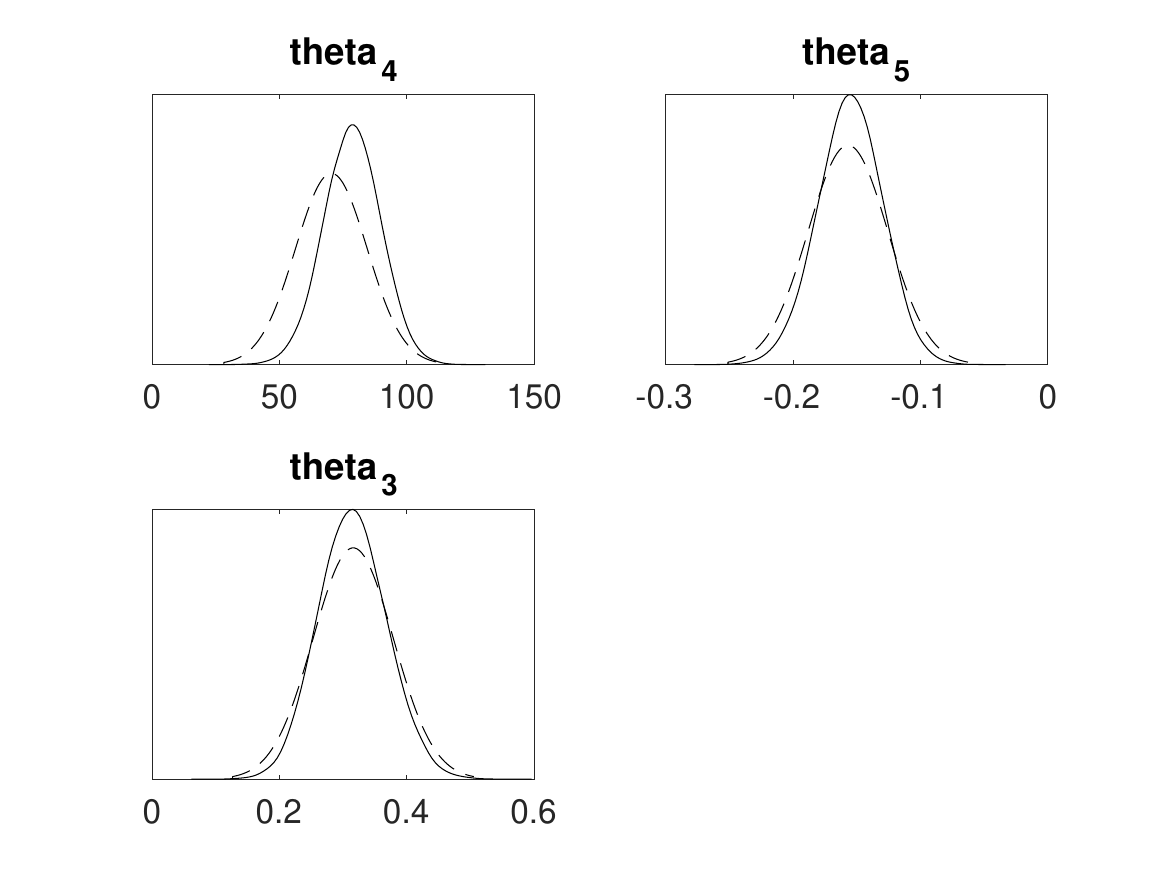}
\includegraphics[width=.33\linewidth]{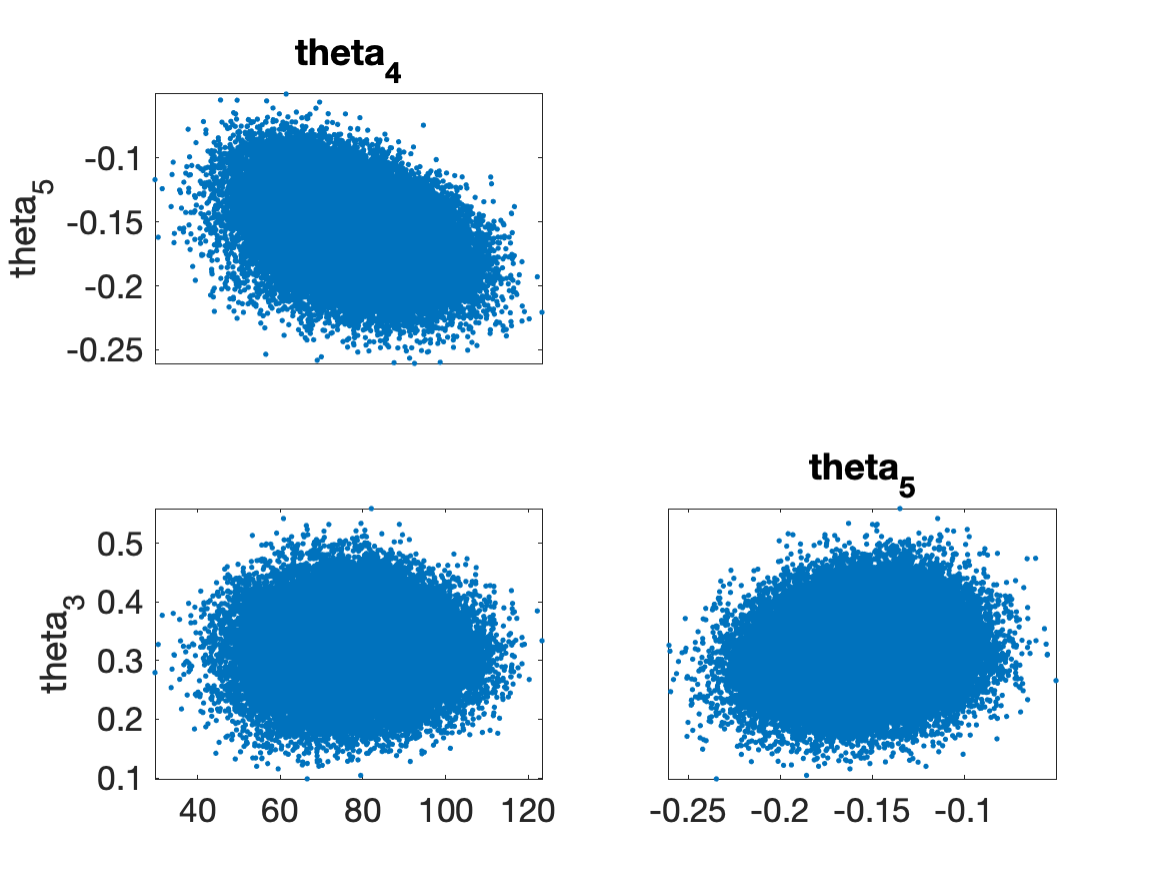} \includegraphics[width=.33\linewidth]{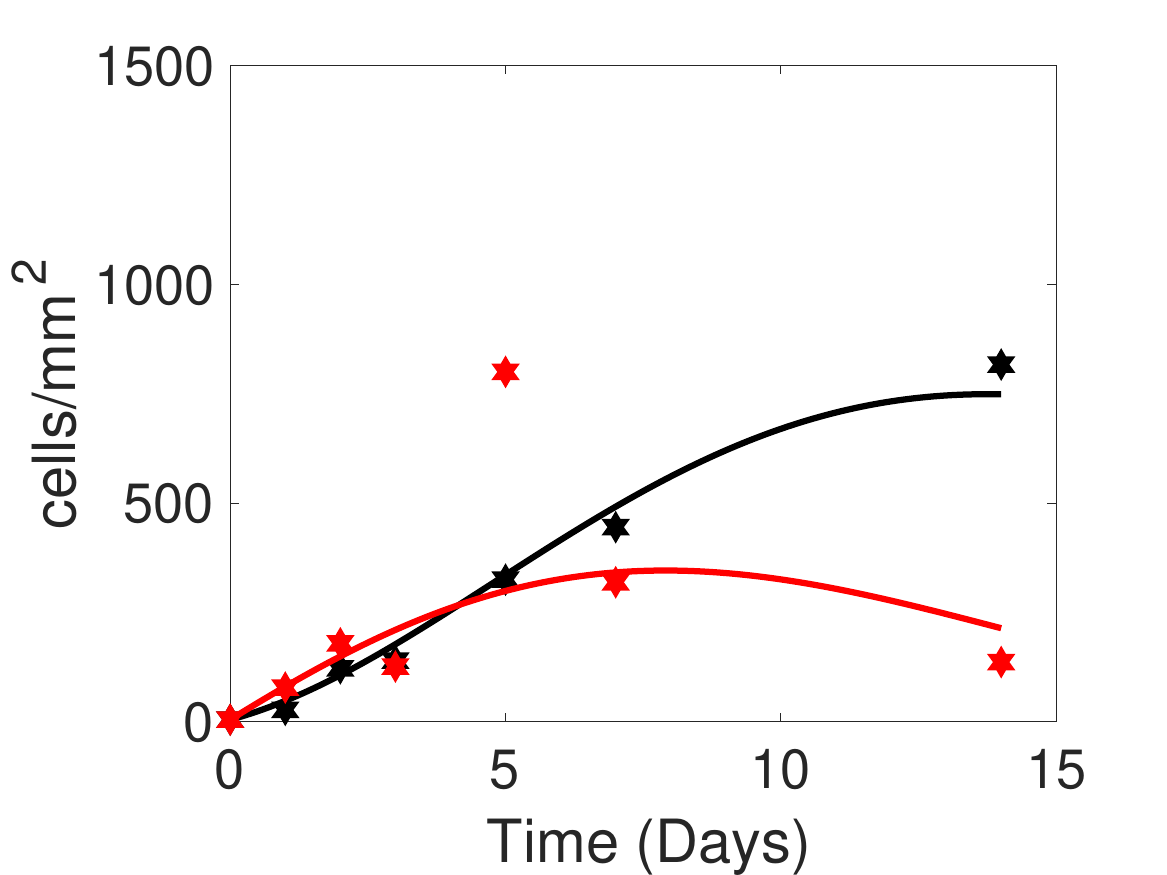} }
\caption{MCMC results for the ODE model in \eqref{eq: SINDyM1}--\eqref{eq: SINDyM2} when estimating the three most sensitive parameters. From left to right: histograms showing the prior (dashed line) and posterior (solid line) distributions of the three estimated parameters; scatter plots showing the correlation between pairs of parameters; and the resulting forward model computed using the posterior means from the estimated posterior distributions.}
\label{fig: DESINDyl01MCMCresults}
\end{figure}

Using the threshold value $\lambda = 0.01$ allows us to balance model error and complexity. Figure~\ref{fig: DESINDyresults01} shows in more detail the results of applying the SINDy method to the M1 and M2 microglial cell count data in Figure~\ref{fig: Data} with a threshold value of $\lambda=0.01$. 
Figure~\ref{fig: DESINDyl01resultsSobol} summarizes the results of applying GSS analysis to the ODE model in \eqref{eq: SINDyM1} and \eqref{eq: SINDyM2}, yielding the three most sensitive parameters: $\theta_4$, $\theta_5$, and $\theta_3$. 
Figure~\ref{fig: DESINDyl01MCMCresults} shows the results of using MCMC with MH sampling to estimate the three most sensitive parameters as determined by GSS. Note that the prior distribution for each estimated parameter in this example is set to be a normal distribution with mean identical to its SINDy point estimate and standard deviation equivalent to $0.2$ times the mean.

\begin{figure}[t!]
\centerline{\includegraphics[width=.35\linewidth]{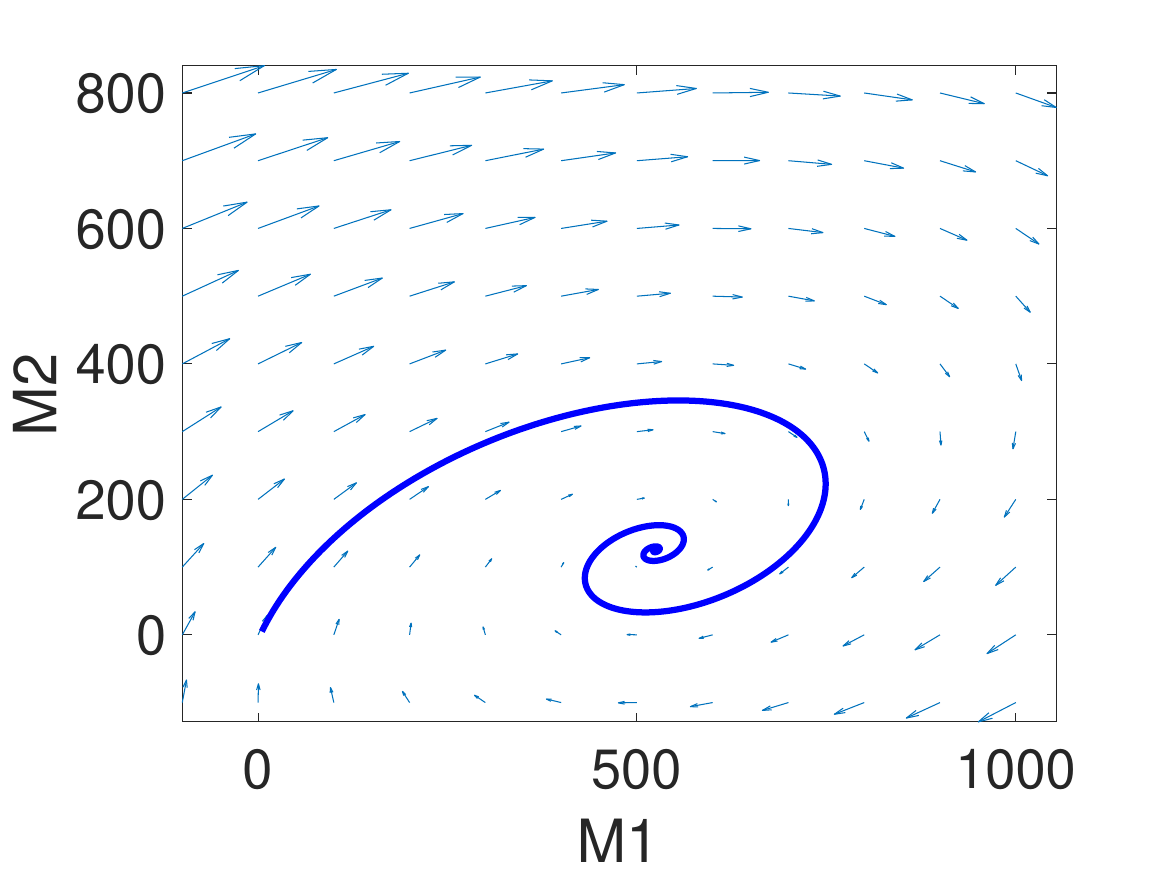}
\includegraphics[width=.35\linewidth]{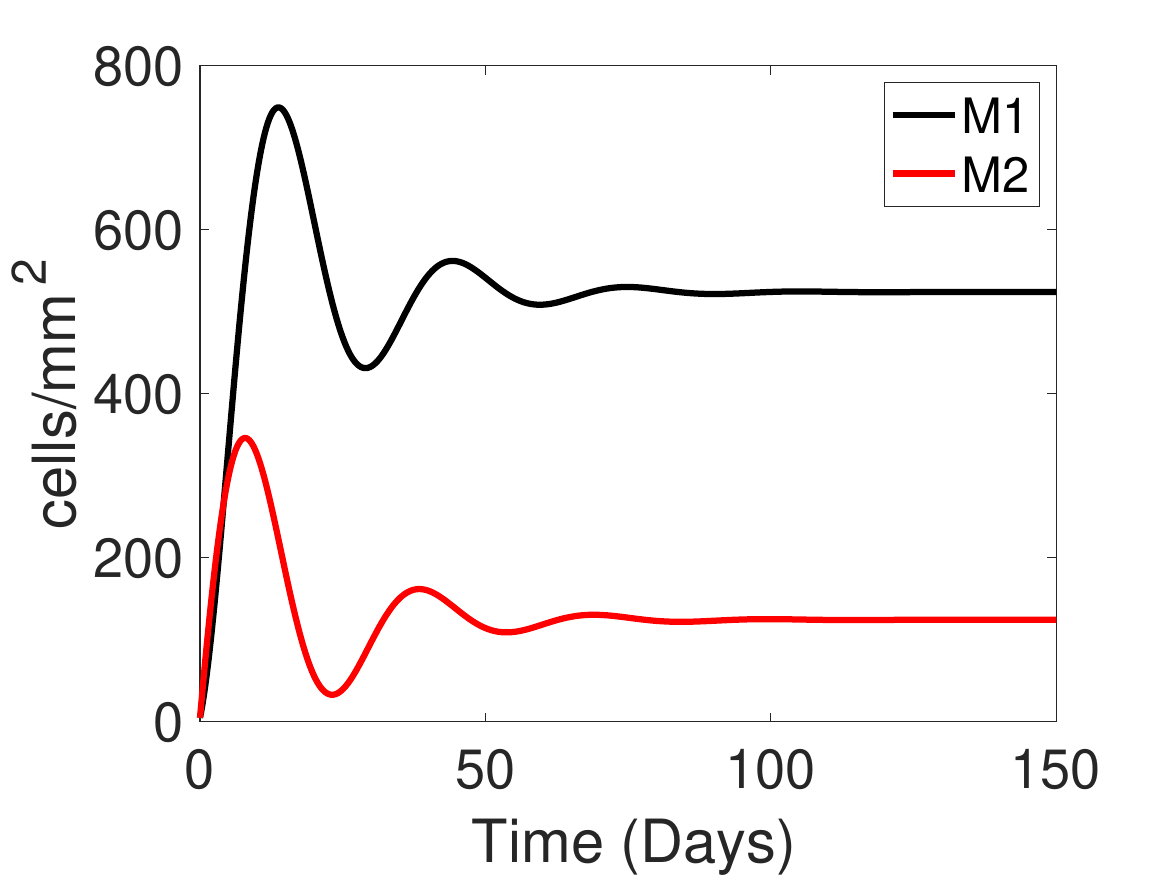} }
\caption{Phase portrait (left) and numerical solution trajectories (right) of the linear ODE system in \eqref{eq:DESINDy_linsys} over the time interval $[0,150]$ days. }
\label{fig: DESINDy_stability}
\end{figure}

As seen in Figure~\ref{fig: DESINDyl01MCMCresults}, the posterior distributions for all three estimated parameters are narrower than their priors, and the mean posterior estimate for $\theta_4$ has shifted more noticeably from its SINDy point estimate. The rightmost panel of Figure~\ref{fig: DESINDyl01MCMCresults} displays the numerical solution to the ODE model in \eqref{eq: SINDyM1}--\eqref{eq: SINDyM2} over the time interval $[0,14]$ days with the values of $\theta_3$, $\theta_4$, and $\theta_5$ set to the mean of their respective posterior distributions and the remaining parameters fixed to their SINDy point estimates. 
Written in matrix form, the corresponding linear system using the MCMC posterior mean estimates for $\theta_3$, $\theta_4$, and $\theta_5$ is given by
\begin{equation}\label{eq:DESINDy_linsys}
    \frac{d}{dt}\left[\begin{array}{c} M1 \\ M2 \end{array}\right] = \left[\begin{array}{c} 32.9650 \\ 78.5130 \end{array}\right] + \underbrace{\left[\begin{array}{cc} -0.1377 & 0.3150 \\ -0.1550 & 0.0217 \end{array}\right]}_{\mathsf{A}} \left[\begin{array}{c} M1 \\ M2 \end{array}\right]
\end{equation}
where the eigenvalues of the coefficient matrix $\mathsf{A}$ are complex conjugate pairs with negative real part, such that the system is stable with sinusoidal oscillations in the long-term solution trajectories. 
Figure~\ref{fig: DESINDy_stability} displays the phase portrait and numerical solution to the linear ODE system in \eqref{eq:DESINDy_linsys} over the span of 150 days.
These plots demonstrate the oscillatory behavior of the solution, which damps over time as the trajectories approach the equilibrium point $(M1, M2)\approx (524, 124)$.

\subsubsection{Forecast Predictions and Biological Implications}

\begin{figure}[t!]
    \centerline{\includegraphics[width=0.44\textwidth]{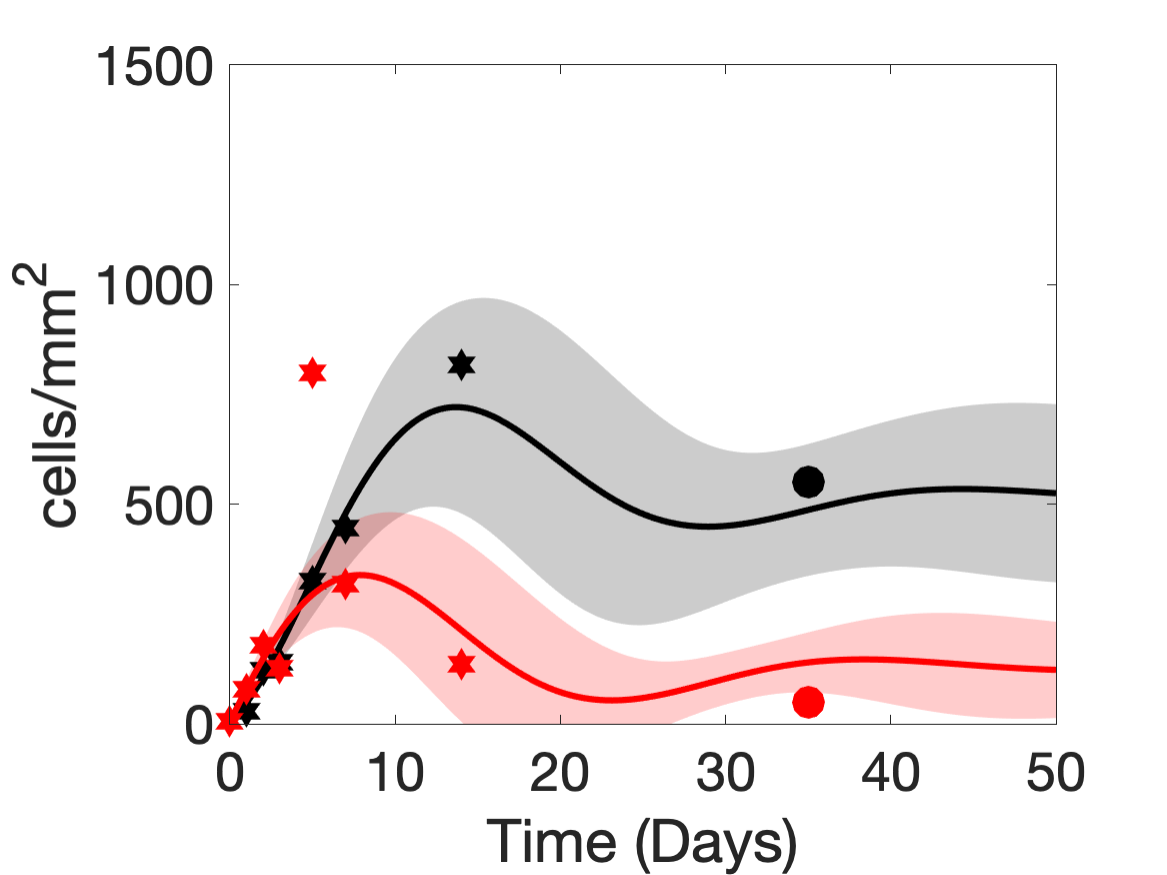}}
    \caption{ODE model forecast predictions with forward propagation of uncertainty. The mean number of M1 (black) and M2 (red) microglial cells are plotted over time with corresponding uncertainty bounds. Black and red stars indicate the microglial cell time series data used within the MCMC algorithm, whereas the black and red solid dots at Day 35 are the number of M1 and M2 microglial cells, respectively, reported in \cite{Suenaga2015}.}
    \label{fig: Exp3pred}
\end{figure}

From a biological perspective, the resulting model in Figure~\ref{fig: DESINDyl01MCMCresults} emphasizes an initial M2 dominance of about four days, followed by an eventual takeover of M1 microglial cells. To make forecast predictions about the number of M1 and M2 microglial cells beyond the 14 days of data that incorporate uncertainty in our estimates, we sample $N = 1,000$ values from the posterior distribution of the MCMC estimated parameters. For each sample, we numerically solve the model using MATLAB's ODE suite \cite{MATLAB_ODE} with parameters set to their sampled values (for $\theta_3$, $\theta_4$, and $\theta_5$) or SINDy point estimates (for $\theta_1$, $\theta_2$, and $\theta_6$) from Day 0 to Day 50 with a time step of $0.1$ days. 
Note that we use \texttt{ode15s} as a safeguard against different parameter combinations potentially causing stiffness.
After repeating this process for all samples, we estimate the mean at each time and use $\pm2$ standard deviations around the mean as uncertainty bounds. Figure~\ref{fig: Exp3pred} shows the resulting forecast predictions over the time interval $[0,50]$ days. 

The uncertainty bounds capture the additional M1 microglial cell count presented in Suenaga et al. (2015)~\cite{Suenaga2015}, while the M2 microglial cell count is right outside the lower uncertainty bound obtained from the SINDy model. Additionally, there are significantly more M1 cells than M2 cells from 14 days on, and the model suggests a persistent inflammatory response. Although experimental studies suggest that neuroinflammation may persist past 14 days, the form of the lingering response is unknown. Our model suggests a decrease followed by an increase for both M1 and M2 cells. After Day 35, intuitively we may expect the cell counts to return to baseline values; however, our results reach an elevated steady state. Therefore, we may lose trust in our results after Day 35 for M1 cells without further data for validation. For M2 cells, we may theorize that an increase at Day 25 should not occur, which would allow the model to better capture the count at Day 35.

Further, we note that the ODE model parameters $\theta_1$, \dots, $\theta_6$ do not have a direct biological interpretation. The SINDy algorithm retains the terms in the basis of functions needed to capture the behavior of the M1 and M2 microglial cell derivatives mathematically and adjusts the parameter values to this effect. 
However, we can hypothesize that, since only linear terms are needed within the equations, the dynamics are not strongly dependent on interactions between the two cell phenotypes.

\subsection{ESINDy+DBN Method}

\begin{figure}[t!]
    \centerline{\includegraphics[width=0.4\textwidth]{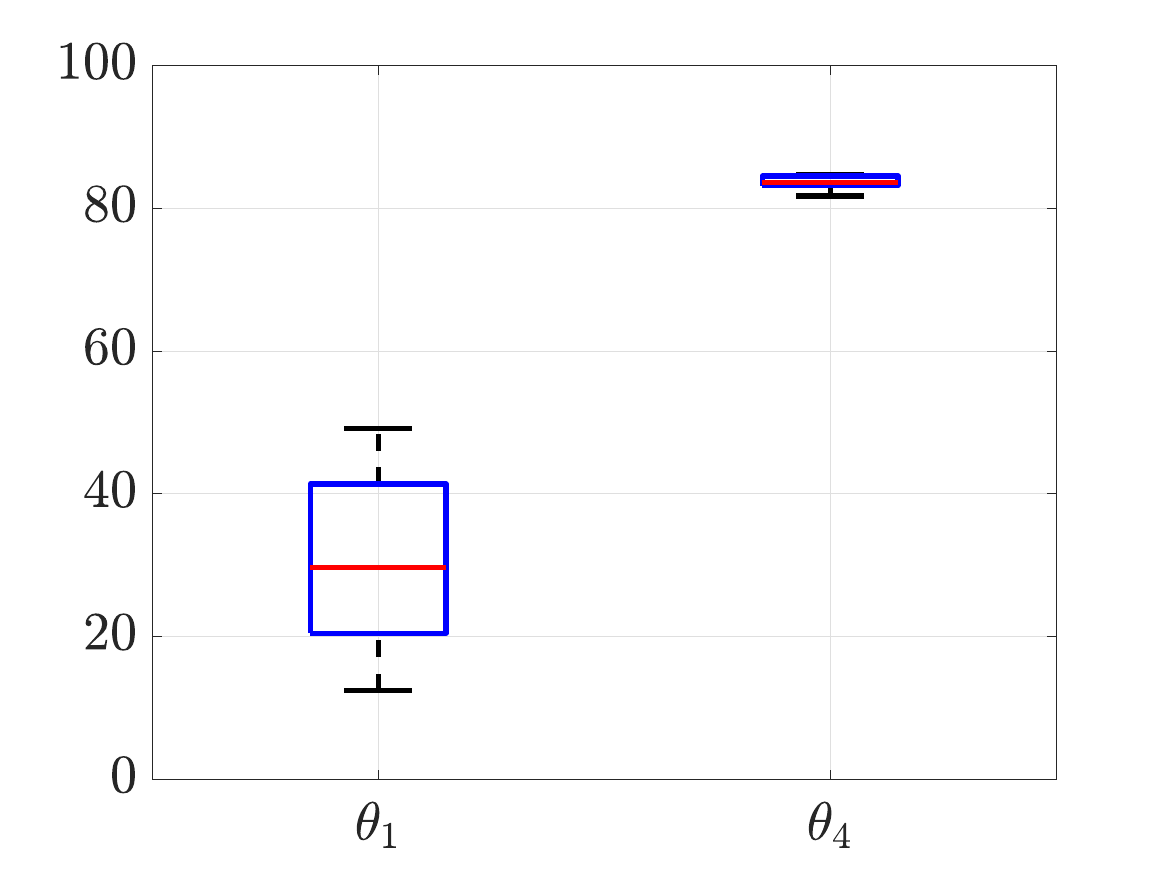} \includegraphics[width=0.4\textwidth]{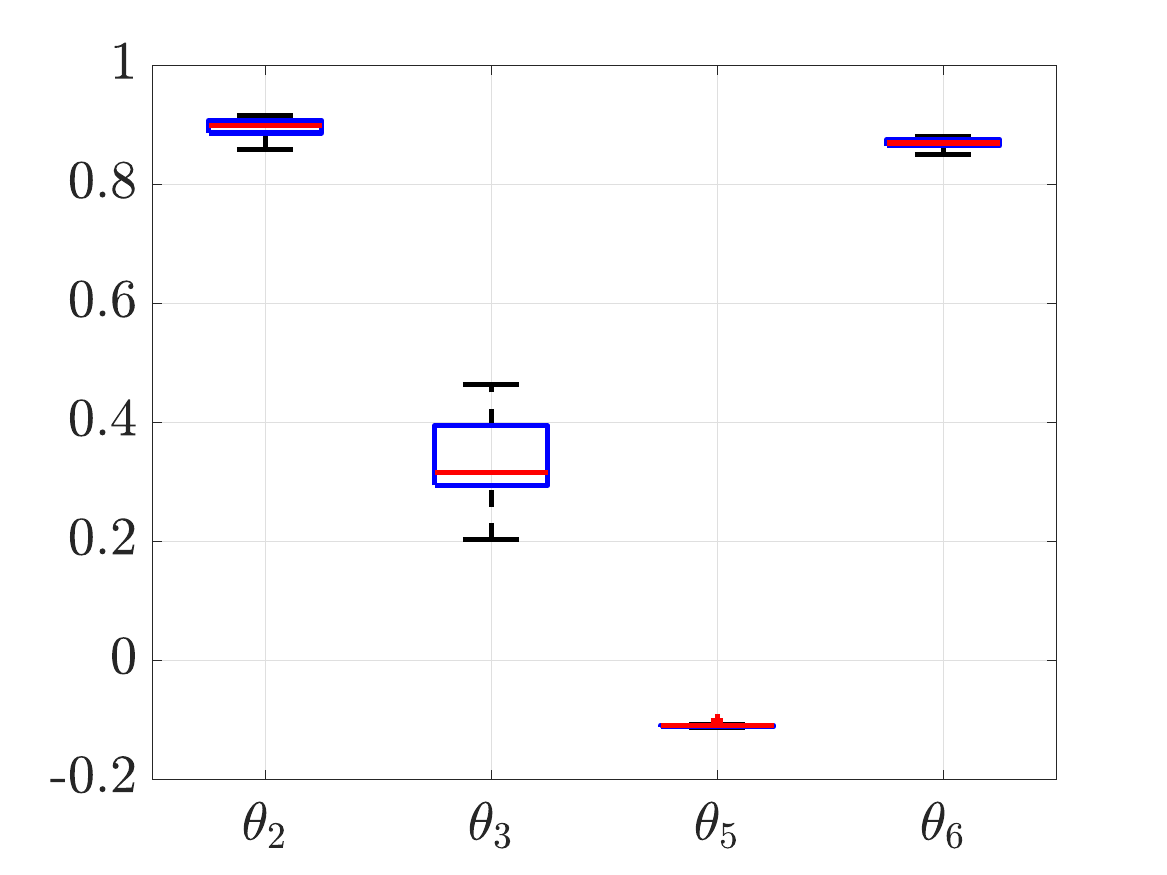}}
    \caption{Range of estimated parameter values across the 10 proxy data sets. The mean value of each coefficient is used to update the discrete-time SINDy model in \eqref{eq: DBNmodel5M1} and \eqref{eq: DBNmodel5M2}.}
    \label{fig: DBN_coeff_boxplots}
\end{figure}

\begin{figure}[t!]
\begin{minipage}{0.45 \textwidth} 
\centering
\includegraphics[width = 0.8 \linewidth]{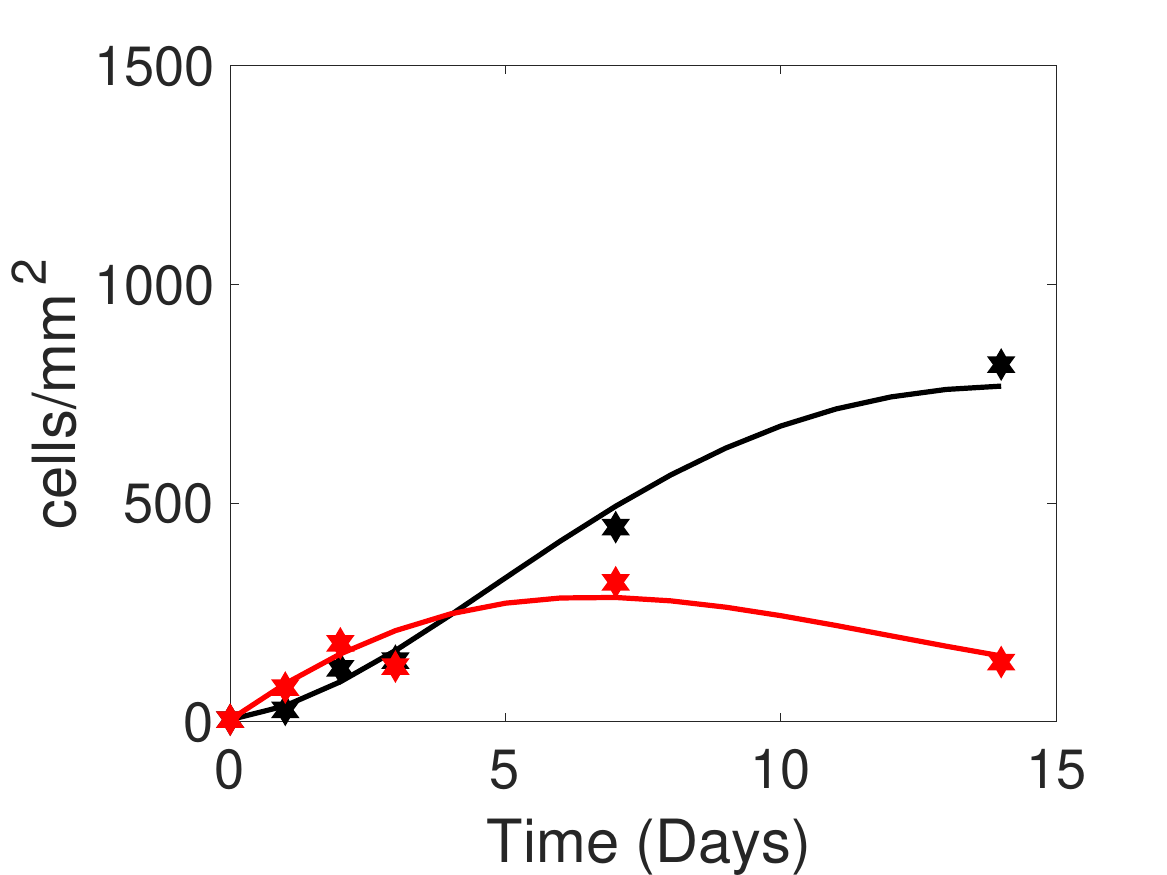}
    \end{minipage}
    \begin{minipage}{0.45 \textwidth}
    \centering
\includegraphics[width = 1.2\linewidth]{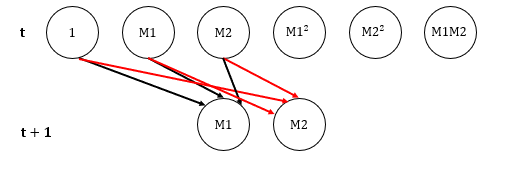}
\end{minipage}
\newline
\begin{minipage}{1 \textwidth} 
    \begin{eqnarray}
        M1_{t+1} &=& \underbrace{30.5432}_{\theta_1} + \underbrace{0.8960}_{\theta_2} M1_t + \underbrace{0.3242}_{\theta_3} M2_t \label{eq: DBNmodel5M1}\\ [0.2 cm]
        M2_{t+1} &=& \underbrace{83.4946}_{\theta_4} + \underbrace{-0.1096}_{\theta_5} M1_t + \underbrace{0.8692}_{\theta_6} M2_t \label{eq: DBNmodel5M2}
    \end{eqnarray}
\end{minipage}
\caption{Results of the averaged model from the ESINDy algorithm. The discrete-time SINDy model has six non-zero parameters, which correspond to the parents of $M1_{t+1}$ and $M2_{t+1}$, respectively. 
}
\label{fig: DBNModel5}
\end{figure}

We apply the process in Section~\ref{subsec: ESINDy_proxydata} to generate 10 proxy data sets for use with ESINDy.
To apply ESINDy in this work, we let $\lambda = 0.01$, $S = 1,000$, and $h = 5$.
For each proxy data set (accounting for different realizations of noise in the simulated measurements), we obtain 4 distinct models from the ESINDy step. In each case, the model which gives 1, $M1_t$, and $M2_t$ as the parents of $M1_{t+1}$ and $M2_{t+1}$ is the most likely model with a probability of around 0.5; this model is illustrated in Figure~\ref{fig: DBNModel5}. Note that, from the DBN perspective, saying ``1 is a parent'' of $M1_{t+1}$ and $M2_{t+1}$ means that there is a constant number of M1 and M2 cells from time $t$ contributing to the total number of M1 and M2 cells at time $t+1$. For the remainder of the results, we focus on one model obtained from averaging the coefficients for 1, $M1_t$, and $M2_t$ in equations for $M1_{t+1}$ and $M2_{t+1}$ from the most likely model from each data set. Figure~\ref{fig: DBN_coeff_boxplots} shows the range of each coefficient value and the mean estimate used in the resulting model, which highlights that $\theta_1$ and $\theta_3$ have the most variance across data sets. The leftmost panel in Figure~\ref{fig: DBNModel5} shows the results of updating the discrete-time SINDy model in \eqref{eq: DBNmodel5M1} and \eqref{eq: DBNmodel5M2} from $t=1$ to $t = 14$ assuming $M1_1 = 5$ and $M2_1 = 5$, with initial conditions taken from the compiled experimental data in Figure~\ref{fig: Data} at Day 0. The rightmost panel of Figure~\ref{fig: DBNModel5} displays the topology of the DBN. The M2 cells are the dominant phenotype until Day 5; after this point, the M1 cells are dominant. This behavior is biologically expected based on experimental studies in the literature and our averaged experimental data set. Further, at Day 14, the M2 cell trajectory appears to be decreasing, while the M1 cell count remains elevated.

\subsubsection{Forecast Predictions and Biological Implications}

\begin{figure}[t!]
\centerline{\includegraphics[width = 0.5\linewidth]{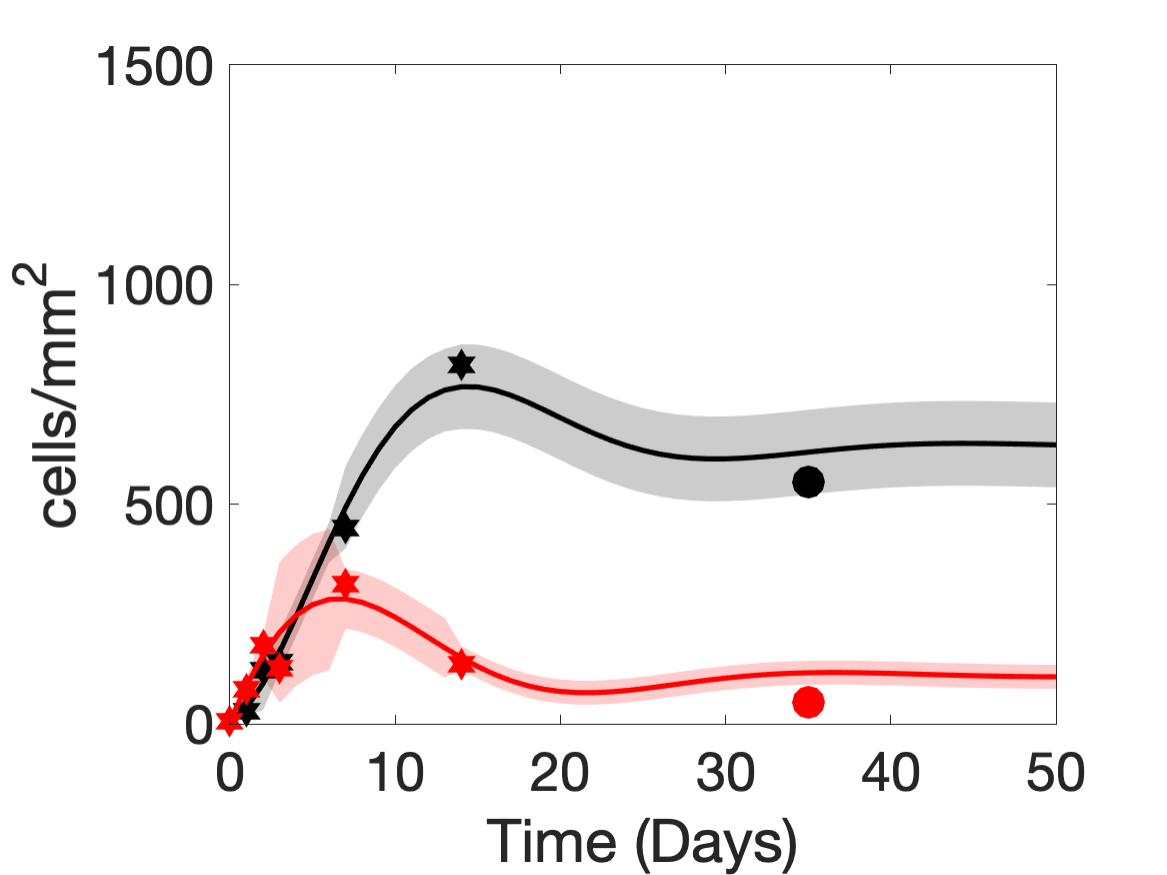}}
    \caption{Forecast predictions with forward propagation of uncertainty of \eqref{eq: DBNmodel5M1prob} and \eqref{eq: DBNmodel5M2prob} over time with a $95\%$ confidence interval. Mean number of M1 (black) and M2 (red) microglial cells over time are plotted with respective confidence intervals. Black and red stars indicate the averaged microglial cell time-series data used to obtain the standard deviation, whereas the black and red solid dots are the number of M1 and M2 microglial cells respectively reported in \cite{Suenaga2015}.}
    \label{fig: DBNmodel5probS}
\end{figure}

The results of the averaged model from the ESINDy algorithm (summarized in Figure~\ref{fig: DBNModel5}) provide the mean for the following conditional probabilities:
\begin{eqnarray}
    p(M1_{t+1}|Pa(M1)_t) &=& \mathcal{N}(30.5432 + 0.8960 M1_t + 0.3242 M2_t, \sigma^2_{M1_{t+1}}) \label{eq: DBNmodel5M1prob} \\ [0.2cm]
    p(M2_{t+1}|Pa(M2)_t) &=& \mathcal{N}(83.4946 - 0.1096 M1_t + 0.8692 M2_t, \sigma^2_{M2_{t+1}}) \label{eq: DBNmodel5M2prob}
\end{eqnarray}
with $\sigma^2_{M1_{t+1}}$ and $\sigma^2_{M2_{t+1}}$ set to the RMSE at each time point as described in Section~\ref{sec: DBN}.
Figure~\ref{fig: DBNmodel5probS} shows the probabilistic predictive modeling results corresponding to the averaged model, plotting the mean of $M1_{t+1}$ and $M2_{t+1}$ along with a $95\%$ confidence interval around each based on the changing variance. Note that the confidence interval is narrow around M1 cells until Day 7, where it widens for the remainder of the time period. The confidence interval for M2 wider earlier in the time series to accommodate data variability before narrowing at Day 14 for the remainder of the time considered.

Similar to the ODE model obtained from the SINDy+MCMC approach, results emphasize an initial dominance of M2 microglial cells of about four to five days, followed by an eventual takeover of M1 microglial cells. The forecast predictions in Figure~\ref{fig: DBNmodel5probS} reflect trends shown in experimental papers of ischemic stroke in various brain areas \cite{Suenaga2015, Rupalla1998, Bodhankar2015}. The additional M1 experimental data point in Suenaga et al. (2015)~\cite{Suenaga2015} lies within the confidence interval of M1 cells, whereas the additional M2 cell measurement at Day 35 lies beneath the confidence interval for M2 cells. There are significantly more M1 cells than M2 cells from 14 days on, and the model suggests a persistent inflammatory response. Our model suggests that the form of the lingering neuroinflammatory response is mostly constant after Day 35. Without additional data beyond this point, we may lose trust in our prediction of M1 cells, since the cell counts do not return to baseline and stay constant at an elevated level. Similarly, we may lose trust in our M2 cell prediction around Day 20, when there is a slight increase in cells which remains consistent and overestimates the data at Day 35. We might expect a small number of cells to remain activated or decrease more towards baseline.

As in the SINDy+MCMC case, our interpretation of the DBN topology may not provide fully relevant biological details about the connections and dynamics between the microglial cells, because different models can yield similar results. Since none of the resulting models obtained from applying the ESINDy+DBN method included nonlinear terms, we can hypothesize that nonlinear interactions are not needed in the formulation of a discrete-time model for M1 and M2 microglial cell counts based on the available data. Biologically, this could provide a computational argument that the M1 and M2 cell interactions are not significantly contributing to their dynamics.

\section{Discussion}
\label{sec: Discussion}

\begin{figure}[t!]
\centerline{\includegraphics[width=0.44\textwidth]{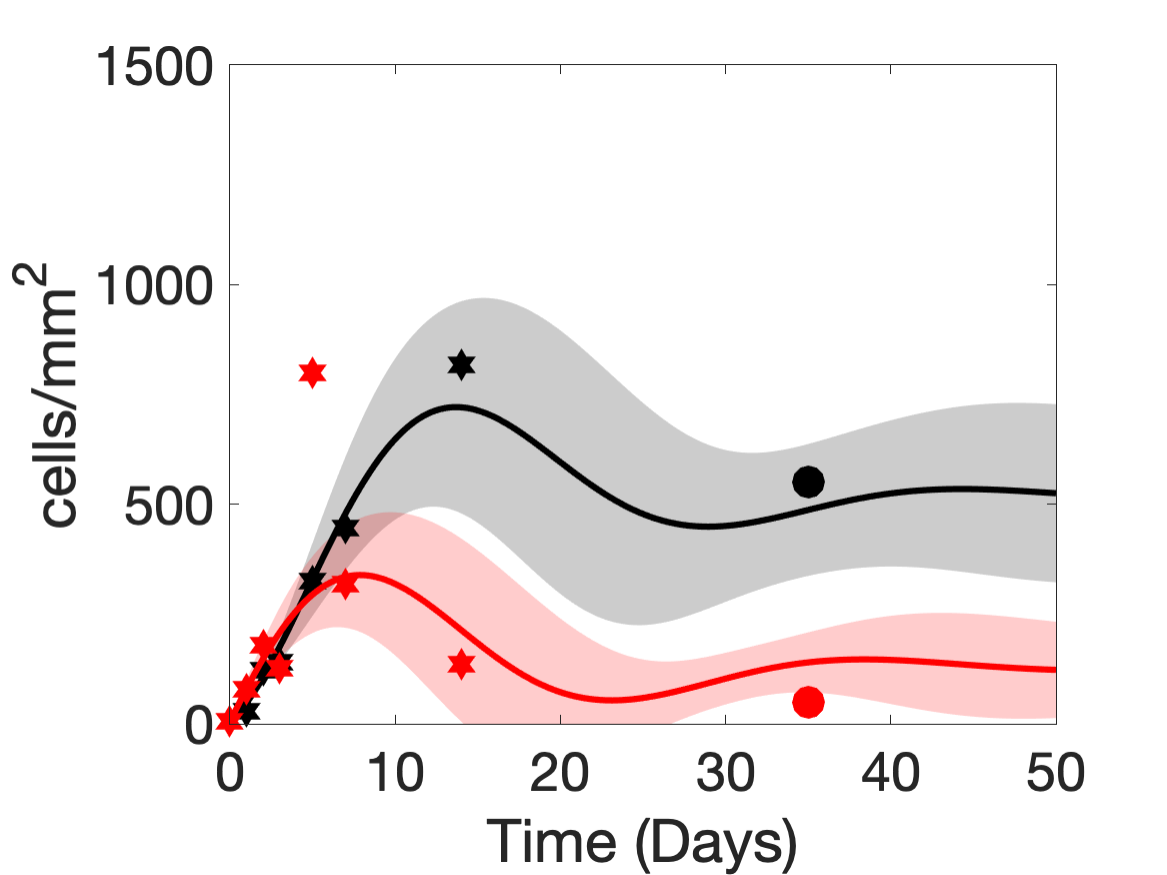} \includegraphics[width=.44\linewidth]{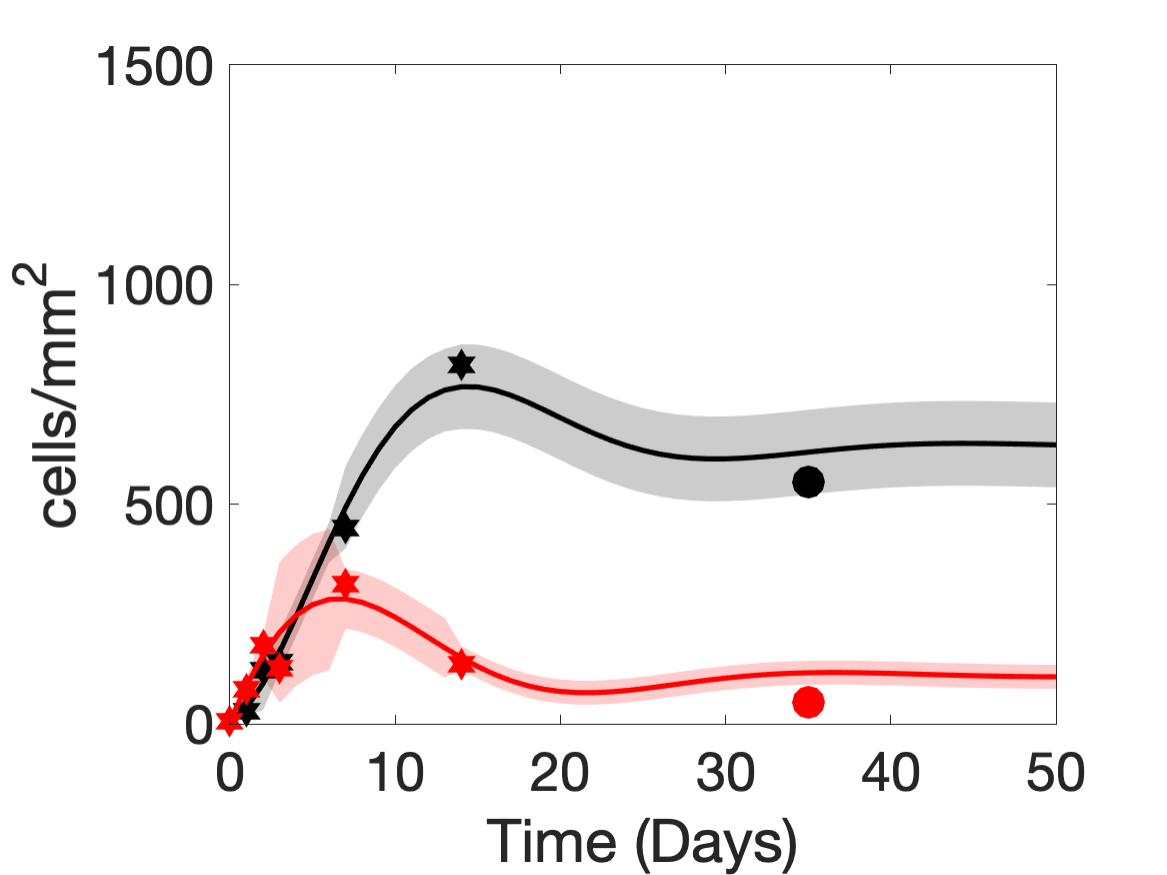}}
\caption{Comparison of forecast predictions with uncertainty quantification for the SINDy+MCMC model (left) and the ESINDy+DBN model (right).}
\label{fig: Compare}
\end{figure}

The data-driven modeling approaches paired with uncertainty quantification techniques in this work yield two new predictive models for microglial cell dynamics following MCAO-induced ischemic stroke in mice.
Figure~\ref{fig: Compare} provides a visual comparison of the forecast predictions with uncertainty for both the SINDy+MCMC and ESINDy+DBN models. Both resulting models highlight the initial dominance of M2 microglial cells; in the SINDy+MCMC model, this dominance is seen for a time period of $3.7$ days, while in the ESINDy+DBN model, this dominance is seen for a time period of four to five days. Due to the discrete nature of the ESINDy+DBN model, we only compute new values of M1 and M2 microglial cells daily, which may contribute to the small discrepancy in the length of M1 dominance between the two models.

Both models also suggest a lingering inflammatory response. The trajectories of the microglial cells from the SINDy+MCMC model show more drastic decreases and increases before reaching a steady state due to the behavior of the resulting ODE system. 
Since the SINDy+MCMC method requires approximating derivatives of the model states from the data, the derivative computation step plays an important role in the resulting model and could be contributing to the more oscillatory dynamics observed in this case. While the ESINDy+DBN method does not rely on derivatives, it does require use of proxy data to augment the experimental measurements at more frequent time points. In future work, we will compare the results in the present work to models obtained by utilizing different data smoothing and augmentation techniques before employing the SINDy steps for each method.
Nonetheless, the combined results of both models suggest that neuroinflammation persists past the initial 14 days. Our resulting model parameters also suggest that M2 cells have a positive effect on the change in M1 cells, possibly indicating that microglial cells switching from the M2 phenotype to the M1 phenotype play an important role in the M1 cell dynamics, while M1 cells have a negative influence on the change in M2 cells, perhaps indicating little to no effect of cells switching from M1 to M2.

The methods presented in this paper provide a data-driven modeling framework to inform our biological understanding of microglial cells and suggest that more experimental studies should be performed to further validate both the biological implications and the long-term dynamics of the cell phenotypes following ischemic stroke. 
While the current models focus solely on microglia, in future work we aim to introduce unmeasured components (such as cytokines) into this modeling framework, using, e.g., time-delay embedding approaches \cite{Bakarji2023}, to analyze how additional cellular components in the affected brain area may contribute to the microglial cell dynamics and overall neuroinflammatory response.
We also plan to compare the proposed methods to an extended data set which directly includes the M1 and M2 measurements at Day 35 from the work of Suenaga et al. (2015)~\cite{Suenaga2015} in order to analyze how additional observations later in the time series affect the resulting models and long-term predictions. 
Note that since our current data set does not include measurements between Days 14 and 35, using polynomial-based approximation techniques to generate proxy data may have limited viability. We will incorporate use of different data augmentation techniques, such as modeling the data with Gaussian processes \cite{Chung2019}, in order to generate a more robust ensemble of proxy measurements for the cell counts in between experimental observations.
For the ESINDy+DBN method, we additionally aim to include computation of the time-dependent variance directly within the ESINDy step to more closely align with the DBN literature. 
Further, while the modeling in this study focuses on cellular-level interactions, our future work will explore integration of these approaches into multiscale models of ischemic stroke that connect the cell level to tissue and patient levels \cite{Li_review2024}.

\section{Conclusion}
\label{sec: Conclusion}

Microglial cells are activated in and recruited to the penumbra during the neuroinflammatory process immediately following the onset of ischemic stroke. In this study, we contribute two new data-driven models to discover, predict, and quantify uncertainty in the dynamics of the two interacting microglial cell phenotypes. We derive the models by combining sparse identification of nonlinear dynamics (SINDy) with Bayesian statistical techniques for robust parameter estimation and uncertainty quantification. Results from both models emphasize an initial M2 dominance followed by a takeover of M1 cells, capture potential long-term dynamics of microglial cells, and suggest a persistent inflammatory response.

\section*{Declaration of Competing Interest}

The authors declare that they have no conflicts of interest.

\section*{Data Availability Statement}

The data that support the findings of this study are available upon reasonable request.

\section*{Acknowledgments}

The authors would like to thank Nils Henninger and Ralph Smith for their valuable comments on this work.


\section*{ORCID iDs}

\noindent Sara Amato: \url{https://orcid.org/0009-0009-5765-2951} \\

\noindent Andrea Arnold: \url{https://orcid.org/0000-0003-3003-882X}



\bibliography{paper_refs.bib}

\begin{thebibliography}{10}

\bibitem{Boche2013}
D.~Boche, V.~H. Perry, and J.~A.~R. Nicoll.
\newblock Activation patterns of microglia and their identification in the
  human brain.
\newblock {\em Neuropathology and Applied Neurobiology}, 39(1):3--18, 2013.

\bibitem{Moulin1985}
D.~E. Moulin, R.~Lo, J.~Chiang, and H.~J. Barnett.
\newblock Prognosis in middle cerebral artery occlusion.
\newblock {\em Stroke}, 16(2):282--284, 1985.

\bibitem{Feske2021}
S.~K. Feske.
\newblock Ischemic stroke.
\newblock {\em The American Journal of Medicine}, 134(12):1457--1464, 2021.

\bibitem{Uzdensky2019}
A.~B. Uzdensky.
\newblock Apoptosis regulation in the penumbra after ischemic stroke:
  expression of pro-and antiapoptotic proteins.
\newblock {\em Apoptosis}, 24:687--702, 2019.

\bibitem{Lee2014}
Y.~Lee, S.~R. Lee, S.~S. Choi, H.~G. Yeo, K.~T. Chang, and H.~J. Lee.
\newblock Therapeutically targeting neuroinflammation and microglia after acute
  ischemic stroke.
\newblock {\em BioMed Research International}, 2014:241--297, 2014.

\bibitem{Guruswamy2017}
R.~Guruswamy and A.~ElAli.
\newblock Complex roles of microglial cells in ischemic stroke pathobiology:
  New insights and future directions.
\newblock {\em International Journal of Molecular Sciences}, 18(3):496--512,
  2017.

\bibitem{Ginsberg2008}
M.~D. Ginsberg.
\newblock Neuroprotection for ischemic stroke: {P}ast, present and future.
\newblock {\em Neuropharmacology}, 55(3):363--389, 2008.

\bibitem{Zhao2017}
S.~C. Zhao, L.~S. Ma, Z.~H. Chu, H.~Xu, W.~Q. Wu, and F.~Liu.
\newblock Regulation of microglial activation in stroke.
\newblock {\em Acta Pharmacologica Sinica}, 38(4):445--458, 2017.

\bibitem{Yenari2010}
M.~A. Yenari, T.~M. Kauppinen, and R.~A. Swanson.
\newblock Microglial activation in stroke: Therapeutic targets.
\newblock {\em Neurotherapeutics}, 7(4):378--391, 2010.

\bibitem{Yang2021}
S.~H. Yang and R.~Liu.
\newblock Four decades of ischemic penumbra and its implication for ischemic
  stroke.
\newblock {\em Translational Stroke Research}, 12:937--945, 2021.

\bibitem{Ma2017}
Y.~Ma, J.~Wang, Y.~Wang, and G.~Y. Yang.
\newblock The biphasic function of microglia in ischemic stroke.
\newblock {\em Progress in Neurobiology}, 157:247--272, 2017.

\bibitem{Hu2012}
X.~Hu, P.~Li, Y.~Guo, H.~Wang, R.~K. Leak, S.~Chen, Y.~Gao, and J.~Chen.
\newblock Microglia/macrophage polarization dynamics reveal novel mechanism of
  injury expansion after focal cerebral ischemia.
\newblock {\em Stroke}, 43(11):3063--3070, 2012.

\bibitem{Li2018}
F.~Li, Q.~Ma, H.~Zhao, R.~Wang, Z.~Tao, Z.~Fan, S.~Zhang, G.~Li, and Y.~Luo.
\newblock L-3-n-{B}utylphthalide reduces ischemic stroke injury and increases
  {M}2 microglial polarization.
\newblock {\em Metabolic Brain Disease}, 33:1995--2003, 2018.

\bibitem{Ma2020}
F.~Ma, P.~Sun, X.~Zhang, M.~H. Hamblin, and K.~Yin.
\newblock Endothelium-targeted deletion of the mi{R}-15a/16-1 cluster
  ameliorates blood-brain barrier dysfunction in ischemic stroke.
\newblock {\em Science {S}ignaling}, 13(626):1--27, 2020.

\bibitem{Wang2017}
R.~Wang, J.~Li, Y.~Duan, Z.~Tao, H.~Zhao, and Y.~Luo.
\newblock Effects of erythropoietin on gliogenesis during cerebral
  ischemic/reperfusion recovery in adult mice.
\newblock {\em {A}ging and {D}isease}, 8(4):410--419, 2017.

\bibitem{Xu2021}
X.~Xu, W.~Gao, L.~Li, J.~Hao, B.~Yang, T.~Wang, L.~Li, X.~Bai, F.~Li, and
  H.~Ren.
\newblock Annexin {A}1 protects against cerebral ischemia--reperfusion injury
  by modulating microglia/macrophage polarization via {FPR}2/{ALX}-dependent
  {AMPK}-m{TOR} pathway.
\newblock {\em {J}ournal of {N}euroinflammation}, 18(1):119--136, 2021.

\bibitem{Li2023}
Y.~Li, Y.~Zhang, Q.~Wang, C.~Wu, G.~Du, and L.~Yang.
\newblock Oleoylethanolamide protects against acute ischemic stroke by
  promoting {PPAR}$\alpha$-mediated microglia/macrophage {M}2 polarization.
\newblock {\em {P}harmaceuticals}, 16(4):621--642, 2023.

\bibitem{Yang2017}
Y.~Yang, H.~Liu, H.~Zhang, Q.~Ye, J.~Wang, B.~Yang, L.~Mao, W.~Zhu, R.~K. Leak,
  and B.~Xiao.
\newblock {ST}2/{IL}-33-dependent microglial response limits acute ischemic
  brain injury.
\newblock {\em {J}ournal of {N}euroscience}, 37(18):4692--4704, 2017.

\bibitem{Murray_book}
J.~D. Murray.
\newblock {\em Mathematical Biology}.
\newblock Springer, New York, 2002.

\bibitem{Calvetti_book}
D.~Calvetti and E.~Somersalo.
\newblock {\em Computational Mathematical Modeling: An Integrated Approach
  Across Scales}.
\newblock SIAM, Philadelphia, 2013.

\bibitem{NODE}
R.~T.~Q. Chen, Y.~Rubanova, J.~Bettencourt, and D.~Duvenaud.
\newblock Neural ordinary differential equations.
\newblock In S.~Bengio, H.~Wallach, H.~Larochelle, K.~Grauman, N.~Cesa-Bianchi,
  and R.~Garnett, editors, {\em Advances in Neural Information Processing
  Systems 31 (NeurIPS 2018)}, page 6571–6583, 2018.

\bibitem{Rackauckas2021UDE}
C.~Rackauckas, Y.~Ma, J.~Martensen, C.~Warner, K.~Zubov, R.~Supekar,
  D.~Skinner, A.~Ramadhan, and A.~Edelman.
\newblock Universal differential equations for scientific machine learning.
\newblock {\em arXiv.org [Preprint]}, 2021.
\newblock \url{https://arxiv.org/abs/2001.04385}.

\bibitem{Brunton2016}
S.~L. Brunton, J.~L. Proctor, and N.~J. Kutz.
\newblock Discovering governing equations from data by sparse identification of
  nonlinear dynamical systems.
\newblock {\em Proceedings of the National Academy of Sciences},
  113(15):3932--3937, 2016.

\bibitem{Zeng2013}
L.~Zeng, Y.~Wang, J.~Liu, L.~Wang, S.~Weng, K.~Chen, E.~F. Domino, and G.~Yang.
\newblock Pro-inflammatory cytokine network in peripheral inflammation response
  to cerebral ischemia.
\newblock {\em Neuroscience Letters}, 548:4--9, 2013.

\bibitem{Lewis2019}
C.~T. Lewis, J.~P.~J. Savarraj, M.~F. McGuire, G.~W. Hergenroeder, H.~A. Choi,
  and R.~S. Kitagawa.
\newblock Elevated inflammation and decreased platelet activity is associated
  with poor outcomes after traumatic brain injury.
\newblock {\em Journal of Clinical Neuroscience}, 70:37--41, 2019.

\bibitem{Azhar2021}
N.~Azhar, R.~A. Namas, K.~Almahmoud, A.~Zaaqoq, O.~A. Malak, D.~Barclay,
  J.~Yin, F.~El-Dehaibi, A.~Abboud, and R.~L. Simmons.
\newblock A putative “chemokine switch” that regulates systemic acute
  inflammation in humans.
\newblock {\em Scientific Reports}, 11(1):1--14, 2021.

\bibitem{Abboud2016}
A.~Abboud, R.~A. Namas, M.~Ramadan, Q.~Mi, K.~Almahmoud, O.~Abdul-Malak,
  N.~Azhar, A.~Zaaqoq, R.~Namas, and D.~A. Barclay.
\newblock Computational analysis supports an early, type 17 cell-associated
  divergence of blunt trauma survival and mortality.
\newblock {\em Critical Care Medicine}, 44(11):1074--1081, 2016.

\bibitem{Alqarni2021}
A.~J. Alqarni, A.~S. Rambely, and I.~Hashim.
\newblock Dynamical simulation of effective stem cell transplantation for
  modulation of microglia responses in stroke treatment.
\newblock {\em Symmetry}, 13(3):404--428, 2021.

\bibitem{Hao2016}
W.~Hao and A.~Friedman.
\newblock Mathematical model on {A}lzheimer's disease.
\newblock {\em {BMC} {S}ystems {B}iology}, 10(1):1--18, 2016.

\bibitem{Shao2013}
H.~Shao, Y.~He, K.~C.~P. Li, and X.~Zhou.
\newblock A system mathematical model of a cell--cell communication network in
  amyotrophic lateral sclerosis.
\newblock {\em {M}olecular {B}io{S}ystems}, 9(3):398--406, 2013.

\bibitem{Vaughan2018}
L.~E. Vaughan, P.~R. Ranganathan, R.~G. Kumar, A.~K. Wagner, and J.~E. Rubin.
\newblock A mathematical model of neuroinflammation in severe clinical
  traumatic brain injury.
\newblock {\em {J}ournal of {N}euroinflammation}, 15(1):1--19, 2018.

\bibitem{Amato2021}
S.~Amato and A.~Arnold.
\newblock Modeling microglia activation and inflammation-based neuroprotectant
  strategies during ischemic stroke.
\newblock {\em Bulletin of Mathematical Biology}, 83:72, 2021.

\bibitem{Amato2024a}
S.~Amato and A.~Arnold.
\newblock A data-informed mathematical model of microglial cell dynamics during
  ischemic stroke in the middle cerebral artery.
\newblock {\em Bulletin of Mathematical Biology}, 87:31, 2025.

\bibitem{Jiang2021}
Y.~X. Jiang, X.~Xiong, S.~Zhang, J.~X. Wang, J.~C. Li, and L.~Du.
\newblock Modeling and prediction of the transmission dynamics of {COVID}-19
  based on the {SIND}y-{LM} method.
\newblock {\em Nonlinear {D}ynamics}, 105(3):2775--2794, 2021.

\bibitem{Mangan2016}
N.~M. Mangan, S.~L. Brunton, J.~L. Proctor, and J.~N. Kutz.
\newblock Inferring biological networks by sparse identification of nonlinear
  dynamics.
\newblock {\em {IEEE} {T}ransactions on {M}olecular, {B}iological and
  {M}ulti-{S}cale {C}ommunications}, 2(1):52--63, 2016.

\bibitem{Sandoz2023}
A.~Sandoz, V.~Ducret, G.~A. Gottwald, G.~Vilmart, and K.~Perron.
\newblock {SIND}y for delay-differential equations: application to model
  bacterial zinc response.
\newblock {\em Proceedings of the {R}oyal {S}ociety {A}}, 479(2269):1--21,
  2023.

\bibitem{Fasel2022}
U.~Fasel, J.~N. Kutz, B.~W. Brunton, and S.~L. Brunton.
\newblock {Ensemble-SINDy}: Robust sparse model discovery in the low-data,
  high-noise limit, with active learning and control.
\newblock {\em Proceedings of the Royal Society A}, 478(2260):1--20, 2022.

\bibitem{Sobol1993}
I.~M. Sobo{\'l}.
\newblock Sensitivity estimates for nonlinear mathematical models.
\newblock {\em Mathematical Modeling and Computational Experiment}, 1:407--420,
  1993.

\bibitem{Smith2013}
R.~C. Smith.
\newblock {\em Uncertainty {Q}uantification: {T}heory, {I}mplementation, and
  {A}pplications}.
\newblock SIAM, Philadelphia, 2013.

\bibitem{Gamboa2013}
F.~Gamboa, A.~Janon, T.~Klein, and A.~Lagnoux.
\newblock Sensitivity indices for multivariate outputs.
\newblock {\em Comptes Rendus. Math{\'e}matique}, 351:307--310, 2013.

\bibitem{Metropolis1953}
N.~Metropolis, A.~W. Rosenbluth, M.~N. Rosenbluth, A.~H. Teller, and E.~Teller.
\newblock Equation of state calculations by fast computing machines.
\newblock {\em J Chem Phys}, 21:1087--1091, 1953.

\bibitem{Hastings1970}
W.~Hastings.
\newblock {M}onte {C}arlo sampling methods using {M}arkov chains and their
  applications.
\newblock {\em Biometrika}, 57:97--109, 1970.

\bibitem{Haario2001}
H.~Haario, E.~Saksman, and J.~Tamminen.
\newblock An adaptive {M}etropolis algorithm.
\newblock {\em Bernoulli}, 7:223--242, 2001.

\bibitem{Haario2006}
H.~Haario, M.~Laine, A.~Mira, and E.~Saksman.
\newblock {DRAM}: efficient adaptive {MCMC}.
\newblock {\em Statistics and Computing}, 16:339--354, 2006.

\bibitem{Robert2004}
C.~P. {R}obert and G.~Casella.
\newblock {\em Monte {C}arlo {S}tatistical {M}ethods}, volume~2.
\newblock Springer, 1999.

\bibitem{Cotter2013}
S.~L. Cotter, G.~O. Roberts, A.~M. Stuart, and D.~White.
\newblock {MCMC} methods for functions: modifying old algorithms to make them
  faster.
\newblock {\em Statistical Science}, 28:424--446, 2013.

\bibitem{Calvetti2023}
D.~Calvetti and E.~Somersalo.
\newblock {\em Bayesian Scientific Computing}.
\newblock Springer Nature, Switzerland, 2023.

\bibitem{Koller2009}
D.~Koller and N.~Friedman.
\newblock {\em Probabilistic Graphical Models: Principles and Techniques}.
\newblock MIT Press, 2009.

\bibitem{AmatoArnold2024_proc}
S.~Amato and A.~Arnold.
\newblock Sparse model identification and prediction of microglial cells during
  ischemic stroke.
\newblock In P.~Nithiarasu and R.~Lohner, editors, {\em 8th International
  Conference on Computational and Mathematical Biomedical Engineering --
  CMBE2024}, pages 322--325, Arlington, VA, USA, 2024.

\bibitem{Suenaga2015}
J.~Suenaga, X.~Hu, H.~Pu, Y.~Shi, S.~H. Hassan, M.~Xu, R.~K. Leak, R.~A.
  Stetler, Y.~Gao, and J.~Chen.
\newblock White matter injury and microglia/macrophage polarization are
  strongly linked with age-related long-term deficits in neurological function
  after stroke.
\newblock {\em {E}xperimental {N}eurology}, 272:109--119, 2015.

\bibitem{AmatoThesis}
S.~Amato.
\newblock {\em Differential Equations and Data-Driven Methods for Modeling
  Microglial Cells During Ischemic Stroke}.
\newblock {PhD} thesis, Worcester Polytechnic Institute, Worcester, MA, 2024.

\bibitem{Sandoz2022}
A.~Sandoz, V.~Ducret, G.~A. Gottwald, G.~Vilmart, and K.~Perron.
\newblock {SIND}y for delay-differential equations: application to model
  bacterial zinc response.
\newblock {\em Proceedings of the {R}oyal {S}ociety {A}}, 479(2269):1--21,
  2023.

\bibitem{Chartrand2011}
R.~Chartrand.
\newblock Numerical differentiation of noisy, nonsmooth data.
\newblock {\em International Scholarly Research Notices}, pages 1--11, 2011.

\bibitem{mcmcstat}
M.~Laine.
\newblock {MCMC toolbox for Matlab}, 2020.
\newblock \url{https://mjlaine.github.io/mcmcstat/}.

\bibitem{Huang2007}
Y.~Huang, J.~Wang, J.~Zhang, M.~Sanchez, and Y.~Wang.
\newblock Bayesian inference of genetic regulatory networks from time series
  microarray data using dynamic {Bayesian} networks.
\newblock {\em Journal of Multimedia}, 2(3):46--56, 2007.

\bibitem{Kim2003}
S.~Y. Kim, S.~Imoto, and S.~Miyano.
\newblock Inferring gene networks from time series microarray data using
  dynamic {Bayesian} networks.
\newblock {\em Briefings in Bioinformatics}, 4(3):228--235, 2003.

\bibitem{Rupalla1998}
K.~Rupalla, P.~R. Allegrini, D.~Sauer, and C.~Wiessner.
\newblock Time course of microglia activation and apoptosis in various brain
  regions after permanent focal cerebral ischemia in mice.
\newblock {\em Acta Neuropathologica}, 96:172--178, 1998.

\bibitem{Bodhankar2015}
S.~Bodhankar, A.~Lapato, Y.~Chen, A.~A. Vandenbark, J.~A. Saugstad, and
  H.~Offner.
\newblock Role for microglia in sex differences after ischemic stroke:
  importance of {M2}.
\newblock {\em Metabolic Brain Disease}, 30:1515--1529, 2015.

\bibitem{Stubbe2013}
T.~Stubbe, F.~Ebner, D.~Richter, O.~R. Engel, J.~Klehmet, G.~Royl, A.~Meisel,
  R.~Nitsch, C.~Meisel, and C.~Brandt.
\newblock Regulatory {T} cells accumulate and proliferate in the ischemic
  hemisphere for up to 30 days after {MCAO}.
\newblock {\em Journal of Cerebral Blood Flow \& Metabolism}, 33(1):37--47,
  2013.

\bibitem{Febinger2015}
H.~Y. Febinger, H.~E. Thomasy, M.~N. Pavlova, K.~M. Ringgold, P.~R. Barf, A.~M.
  George, J.~N. Grillo, A.~D. Bachstetter, J.~A. Garcia, and A.~E. Cardona.
\newblock Time-dependent effects of {CX3CR1} in a mouse model of mild traumatic
  brain injury.
\newblock {\em Journal of Neuroinflammation}, 12:1--16, 2015.

\bibitem{Younger2019}
D.~Younger, M.~Murugan, K.~V.~R. Rama, L.~Wu, and N.~Chandra.
\newblock Microglia receptors in animal models of traumatic brain injury.
\newblock {\em Molecular Neurobiology}, 56:5202--5228, 2019.

\bibitem{Donat2017}
C.~K. Donat, G.~Scott, S.~M. Gentleman, and M.~Sastre.
\newblock Microglial activation in traumatic brain injury.
\newblock {\em Frontiers in Aging Neuroscience}, 9:208--229, 2017.

\bibitem{MATLAB_ODE}
L.~F. Shampine and M.~W. Reichelt.
\newblock {The MATLAB ODE Suite}.
\newblock {\em SIAM Journal on Scientific Computing}, 18:1--22, 1997.

\bibitem{Bakarji2023}
J.~Bakarji, K.~Champion, J.~N. Kutz, and S.~L. Brunton.
\newblock Discovering governing equations from partial measurements with deep
  delay autoencoders.
\newblock {\em Proceedings of the Royal Society A}, 479:20230422, 2023.

\bibitem{Chung2019}
M.~Chung, M.~Binois, R.~B. Gramacy, J.~M. Bardsley, D.~J. Moquin, A.~P. Smith,
  and A.~M. Smith.
\newblock Parameter and uncertainty estimation for dynamical systems using
  surrogate stochastic processes.
\newblock {\em SIAM Journal on Scientific Computing}, 41:A2212--A2238, 2019.

\bibitem{Li_review2024}
G.~Li, Y.~Zhao, W.~Ma, Y.~Gao, and C.~Zhao.
\newblock Systems-level computational modeling in ischemic stroke: from cells
  to patients.
\newblock {\em Frontiers in Physiology}, 15:1394740, 2024.

\end{thebibliography}


\end{document}